\documentclass[structabstract]{aa}
\usepackage{graphicx}
\usepackage{txfonts}

\usepackage{natbib,url}

\usepackage{xcolor}

\begin{document}

\title{Evolution and motions of magnetic fragments during the active region formation and decay: A statistical study}

\author {Michal \v{S}vanda\inst{1,2}
       \and
       Michal Sobotka\inst{2}
       \and
        Lucia Mravcov\'a\inst{2,1}
        \and
        Tatiana V\'ybo\v{s}\v{t}okov\'a\inst{3}
        }

\offprints{M.~\v{S}vanda, 
\email{svanda@sirrah.troja.mff.cuni.cz}}

\institute
    {
     Charles University, Astronomical Institute, V~Hole\v{s}ovi\v{c}k\'{a}ch 2, CZ-18000, Prague 8, Czech Republic
      \and
      Astronomical Institute of the Czech Academy of Sciences, Fri\v{c}ova 298, CZ-25165 Ond\v{r}ejov, Czech Republic
     \and
     Charles University, Department of Surface and Plasma Science, V~Hole\v{s}ovi\v{c}k\'{a}ch 2, CZ-18000, Prague 8, Czech Republic
     }

\date{Received date / Accepted date }
\titlerunning{Evolution of magnetic fragments in active regions}
\authorrunning{\v{S}vanda et al.}

\abstract
{The evolution of solar active regions is still not fully understood. The growth and decay of active regions have mostly been studied in case-by-case studies.}
{Instead of studying the evolution of active regions case by case, we performed a large-scale statistical study to find indications for the statistically most frequent scenario.}
{We studied a large sample of active regions recorded by the Helioseismic and Magnetic Imager instrument. The sample was split into two groups: forming (367 members) and decaying (679 members) active regions. We tracked individual dark features (i.e. those that are assumed to be intensity counterparts of magnetised fragments from small objects to proper sunspots) and followed their evolution. We investigated the statistically most often locations of fragment merging and splitting as well as their properties.}
{Our results confirm that statistically, sunspots form by merging events of smaller fragments. The coalescence process is driven by turbulent diffusion in a process similar to random-walk, where supergranular flows seem to play an important role. The number of appearing fragments does not seem to significantly correlate with the number of sunspots formed. The formation seems to be consistent with the magnetic field accumulation. Statistically, the merging occurs most often between a large and a much smaller object. The decay of the active region seems to take place preferably by a process similar to the erosion. }
{}


\maketitle

%
\section{Introduction}
Strong local magnetic fields forming active regions are the main manifestations of solar magnetic activity. The formation, evolution, and structure of these active regions have been studied for more than four centuries; however, they are still not completely understood. Reliable information about the structure of active regions beneath the surface does not exist \citep{2009arXiv0912.4982M}. Although there were attempts to use local helioseismology to infer the depth structure of sunspots \citep[e.g.][]{2001ApJ...557..384Z}, they were largely questioned recently \citep[e.g.][]{2009SSRv..144..249G}. Thus all of our useful knowledge about the 3D structure of, for example, sunspots and their surroundings comes from theoretical models \citep[e.g.][]{2009ApJ...691..640R} or topological studies \citep[e.g.][]{2013ApJ...764L...3C}. 

The current paradigm is that solar magnetic fields are generated by dynamo action deep in the convection zone. When the magnetic field is sufficiently strong, it becomes buoyant and emerges in the form of toroidal flux ropes oriented in the east-west direction, forming bipolar active regions on the surface. The emergence of the flux alternates the mode of convection, which is mostly represented by granular cells at the surface. This was studied by \cite{2007A&A...467..703C}, for example, in a numerical simulation. They concluded that due to the high Mach numbers found in convective
downflows, it is virtually impossible for buoyant flux tubes to rise unimpeded by the convection.

As a result of the very vigorous convection near the solar surface, the morphology of emerging flux regions will be highly structured by the near-surface convection. A general overview of the evolution of active regions was delivered by \cite{2015LRSP...12....1V}, where the authors, in general, describe the processes considered during the active regions' lifetime starting from the pre-emergence phases and ending by fading the enhanced network to join the background magnetic field. Despite the very thorough description, the authors conclude that not all the phases of the active regions' evolution are understood sufficiently. 

Unfortunately, lately, specific phases of the evolution, the morphology of active regions, as well as large sunspots namely have not drawn the attention  required. For instance, according to the classical paper by \cite{1985SoPh..100..397Z}, the sunspots form by coalescence of a few larger fragments, which are themselves made of typically two-three fragments. According to \cite{1987SoPh..112...49G}, these fragments retain their identity during the whole evolution of sunspots, even though they change their size, shape, and position. When the sunspot starts to decay, its break-up starts at the boundaries of those fragments, and light bridges usually form there. Some of these fragments survive the sunspot's break-up and turn in the long-lived spots of simpler magnetic configuration of late Z\"urich type. Such a picture deserves confirmation in a statistical sense using state-of-the-art data. 

On the other hand, decaying sunspots with penumbra are usually surrounded by a moat -- a region around the sunspot, where the flows prevail radially out from the sunspot \citep{1969SoPh....9..347S}. Usually, moats are present on those sides of the sunspot where a penumbra also exists. Within the moat, high-resolution observations showed that small magnetic elements (termed moving magnetic features, MMFs) rush away from the spot \citep{2005ApJ...635..659H}. At the outer boundary of the moat, MMFs disappear or merge into the magnetic network. The sunspots' decay via MMFs seems to represent a different scenario than the decay by large fragments. The formations or decays of sunspots were typically studied on a case-by-case basis \citep[e.g.][]{2017IAUS..328..127C}, mostly using campaign high-resolution data \citep[e.g.][]{2020IAUS..354...53K}. These processes were also targets of numerical simulations. 

State-of-the-art numerical simulations show the sunspot evolution from the
modellers point of view. For example, \cite{2010ApJ...720..233C} showed that when the bipolar active region emerges, first, small-scale magnetic elements appear at the surface. They gradually coalesce into larger magnetic concentrations, which eventually results in the formation of a pair of opposite polarity spots. The whole active region evolution was simulated by \cite{2014ApJ...785...90R}; the description includes the decay of the active region, which was driven by the flows from the sub-surface. An almost field-free plasma intruded the magnetic field and when these intrusions reached the photosphere, the spot fragmented. Dispersal of the flux from the simulated sunspot was consistent with the decay by turbulent diffusion. 

Similarly, \cite{2017ApJ...846..149C} studied the emergence of a magnetic flux bundle from the convection zone to the corona. The numerical simulation among others resulted in the synthetic continuum intensity images that were equivalent to high-resolution observations. The simulation shows that the magnetic flux bundle rose as a coherent structure throughout the upper convection zone, except for the uppermost several megametres, where the bundle fragmented. These fragments consisted of magnetic flux tubes emerging individually at the surface. The synthetic white-light images showed that first, small granular-sized magnetic elements appeared at the surface and they gradually coalesced to pores. As the emergence continued, the pores merged to sunspots. 

It would seem that coalescence of small-scale magnetic features with the formation of larger structures is dominant during the magnetic flux emergence (this process is referred to as inverse turbulent cascading, where the energy transfers from small scales to larger scales). The full-scale inverse cascade was observationally reported for the first time by \cite{2008SoPh..248..311H} when studying a sequence of magnetograms of a particularly fast emerging active region NOAA~10488. 

The emergence takes place in a very turbulent medium in the upper solar convection zone, so the opposite process occurs at the same time. This process is referred to as the direct cascade when the magnetic structures naturally fragment to smaller ones. The interplay of both cascades during the emergence of several active regions was confirmed by \cite{2019SoPh..294..102K}. They conclude that most of the time the energy grew at all scales. The study could not state that the inverse cascade was a dominant process during the formation of an active region. Although coalescence of small magnetic elements into larger pores and sunspots was observed, in terms of energy contribution to the active region the emergence of large-scale structures was more important. 

The availability of the synoptic observations with public available archives increased the number of captured sunspot formations and decays. In the modern-most synoptic observations by the Helioseismic and Magnetic Imager \citep[HMI;][]{2012SoPh..275..207S} on board of the Solar Dynamics Observatory \citep[][]{2012SoPh..275....3P}, the spatial sampling of 0.5" is sufficient enough to resolve small magnetic features from which sunspots are supposed to be formed by coalescence according to the scenario of \citeauthor{1987SoPh..112...49G}. The long-term coverage of over 10 years with a duty cycle close to 100\% allows for one to select suitable active regions recorded during these critical phases of their evolution. Instead of studying their evolution case by case, we performed a large-scale statistical study to find indications for a general scenario. This general or most frequent scenario does not preclude other different possibilities. Nevertheless, these different possibilities are statistically less likely.

\section{Data}
From the HMI archive of pseudocontinuum intensitygrams (data series {\tt hmi.Ic\_45s}), we selected sunspots having the following properties. 

For the forming active regions we required that (1) an active region emerged not farther than about 60 degrees from the disc centre and (2) it survived within this distance for at least 3 days. For further analysis, we considered all of the days when the active region was within the central meridian distance (CMD) of 60 degrees. Additionally, we also considered 2 days before the emergence regardless of the CMD of the prospective active region location. In total, a sample of 367 members was identified. 

For decaying active regions we required that (1) an active region decayed not farther than about 60 degrees from the disc centre and (2) before the final decay, it was observable for at least 3 days within the distance of 60 degrees from the central meridian. For further analysis, we considered all of the days when the active region was within the CMD of 60 degrees. Furthermore, we also considered one additional day after the final decay regardless of the CMD of the past active region's location. In total, 679 members were identified in the data archives. 

The selection of the appropriate active regions was done manually closely cooperating with NASA's SolarMonitor\footnote{\url{www.solarmonitor.org}}. The time range covered was between May 2010 and August 2018. The dismemberment into a forming or decaying group was done subjectively following the simplest possible criteria: When the active region emerged within a CMD of 60 degrees, it was considered for the group of forming active regions. Similarly, when the active region finally decayed (i.e. no sunspots were observed on the following days) within a CMD of 60 degrees, it was considered for the group of decaying active regions. The days before the emergence or after the final decay, respectively, were considered as the safety margins to make sure we did not miss the appearance of possible small ephemeral objects on those days. 

For each group, we tracked the datacube using the Carrington rotation and selected the field of view having 768$\times 768$ pixels centred on the AR coordinates given by SolarMonitor. The datacube consisted of two spatial coordinates (one corresponding to the zonal east-west direction and the other to the meridional south-north direction) and one time coordinate. The frames were Postel projected with a pixel size at the centre of the field of view of 0.0301 heliographic degrees, which corresponds to 0.366 Mm. The frame cadence was 12 minutes, which was a trade-off between the data volume and the time resolution. The tracking and mapping were done using the JSOC\footnote{Joint Science Operation Centre, \url{jsoc.stanford.edu}} tools. 

The missing frames were linearly interpolated so that the datacube had a regular sampling in time. For each considered datacube, we also created a visual movie for the inspection of the evolution by eye. 

\section{Methods}

\subsection{Fragment detection}
\label{sect:object_detect}
Henceforth, the terms `object' and `fragment' indicate the same quantity, that is a dark object detected by the pipeline described in this section. The pipeline is based on the object tracking algorithm developed by
\cite{1997A&A...328..682S} for tracking small-scale features in sunspots.

In the first step, continuum-intensity frames that include the edge of the solar disc, appearing at the beginning and/or end of the time series, are discarded. Then, the centre-to-limb intensity gradient is removed to obtain flattened frames with photospheric intensity $I_{\rm ph}$ normalised to unity. The normalisation value is obtained by the flattened field of view averaging with a clearly dominant contribution of the quiet photosphere. Simple image segmentation is applied to these frames, defining objects as groups of pixels with intensities of $I_{\rm ph} < 0.9$. Side-by-side neighbouring pixels form the same object. Single-pixel objects are considered as noise and discarded. The objects retain their original intensity, while the intensity in the rest of the field of view is set to zero.

In the second step, evolutionary histories of objects are tracked in a series of segmented frames. An object continues its history if at least one of its pixels coincides in position with an object's predecessor in the preceding frame. There are four possibilities for the object-related event: the object appears, it may fade out, it may merge with another object, or it may split in two. The rules applied for merging and splitting are described in Section~\ref{sect:splitmerge}. The code records the area-averaged intensity, the coordinates of the centre of gravity, and the number of pixels (area) of each object for each time instant (frame) of its history. In the tracking results, objects with time-averaged areas smaller than nine~pixels, that is to say 1.2~Mm$^2$ which is comparable to a large granule, are considered noise and discarded.

In the third step, the tracking results are used to map events of merging and splitting. For the instant of birth or death of a tracked object, the code looks for a preceding or successive neighbour object, from or to which the tracked object can split or merge (see Section~\ref{sect:splitmerge}). When such an object is found, information about the two objects, time, and the type of event (split or merge) is recorded.

\subsection{Sunspot detection}
\label{sect:sunspot_detect}
The object detection pipeline detects all objects that have an  intensity below a given threshold. That also includes sunspots.
It is not straightforward to distinguish sunspots from other objects (fragments) using just a simple criterion. On the other hand, to deal with the science goals of our study, the need for a reference series containing only detected sunspots is perspicuous. 

To detect only sunspots, we chose to follow the methodology already published in the literature. We used an object identification algorithm based on the mathematical morphology, following the procedure described by \cite{2009SoPh..260....5W}, and coded using Python's Scikit-Image library. 

The continuum datacube was processed frame-by-frame. For each frame, we applied the median filter (with a boxcar of 20 px) to remove the small-scale noise such as the cosmic-ray hits or very small features. 
Then we applied a morphological top-hat transform \citep{2003SPIE...D} to image $I$. Top-hat transforms are used for various image processing tasks, such as feature extraction, background equalisation, image enhancement, and others. We used the white top-hat transform, which is defined as the difference between the input image and its opening by some structuring element. The structuring element must be larger than the features in order to be detected. As a structuring element $A$, we used the circle having a diameter of 196~px, which corresponds to about 72~Mm. 

The white top-hat transform is then defined by
\begin{equation}
T_{w}(I)=I-I\circ A,
\end{equation}
where $\circ$ indicates the morphological opening operation, which corresponds to an erosion followed by a dilation. We note that the white top-hat transform is designed to extract brighter features, so in order to extract dark sunspots, one has to invert image $I$ first. A nice example of the individual steps of the procedure is given in Fig.~1 in \cite{2009SoPh..260....5W}.  

All pixels of $T_{w}(I)$ larger than 6000~DN/s (an experimental threshold) are considered as sunspots\footnote{A short note regarding the units: In Section~\ref{sect:object_detect} the algorithm utilised flattened intensity images, which were normalised to the quiet-Sun intensity. In the case of Section~\ref{sect:sunspot_detect}, we kept the HMI {\tt I$_c$} units `data number per second. Sunspots areas segmented by thresholding the difference between the image and its background obtained by the application of the opening operator exceeding the threshold. Hence the value of the threshold keeps the data units.}. Automatic labelling of continuous regions was performed to identify individual sunspots. Due to the median filter, only sunspots larger than a few megametres were detected. Small pores avoid the detection but they were retained as fragments in the fragment detection pipeline.

The thresholds of both detection pipelines were set such that any feature at least 10\% darker than the quiet photosphere was identified as a separate object. That is, umbrae and penumbrae fall below the thresholds. In the case of evolved sunspot, it is identified as one compact object including the penumbra surrounding the umbra as long as they are not completely separated by a bright structure corresponding to the quiet-Sun photosphere brightness. The very small and short-lived features are considered to be noise and removed by the application of the size criteria. This application certainly also removes some real, short-lived small objects. We believe that this omission does not affect our analysis significantly and that the benefits of discarding the noise outweigh the loss of a certain object population. 

In about 10\% of the active regions in our sample, no proper sunspots were detected by the sunspot detection pipeline. These active regions are only populated by small fragments. Naturally, in those active regions, the number of detected objects is also much lower. In total, the objects contained in the active regions without sunspots represent 1.1\% of the total. Therefore we consider any possible influences of these spotless active regions to the results and conclusions negligible. 

\subsection{Merging and splitting centres of fragments}
\label{sect:splitmerge}
The object tracking algorithm allows one to build a time-space trajectory and identify possible merging or splitting events.
The merging event is considered if a single object in a given frame overlaps the pixels of two objects (at least one pixel each) in the preceding frame. It means that two objects converged to a common location. The follow-up object is considered to be a merger and keeps the label of the older merging object. 
\begin{figure*}
    \centering
    \includegraphics[width=0.8\textwidth]{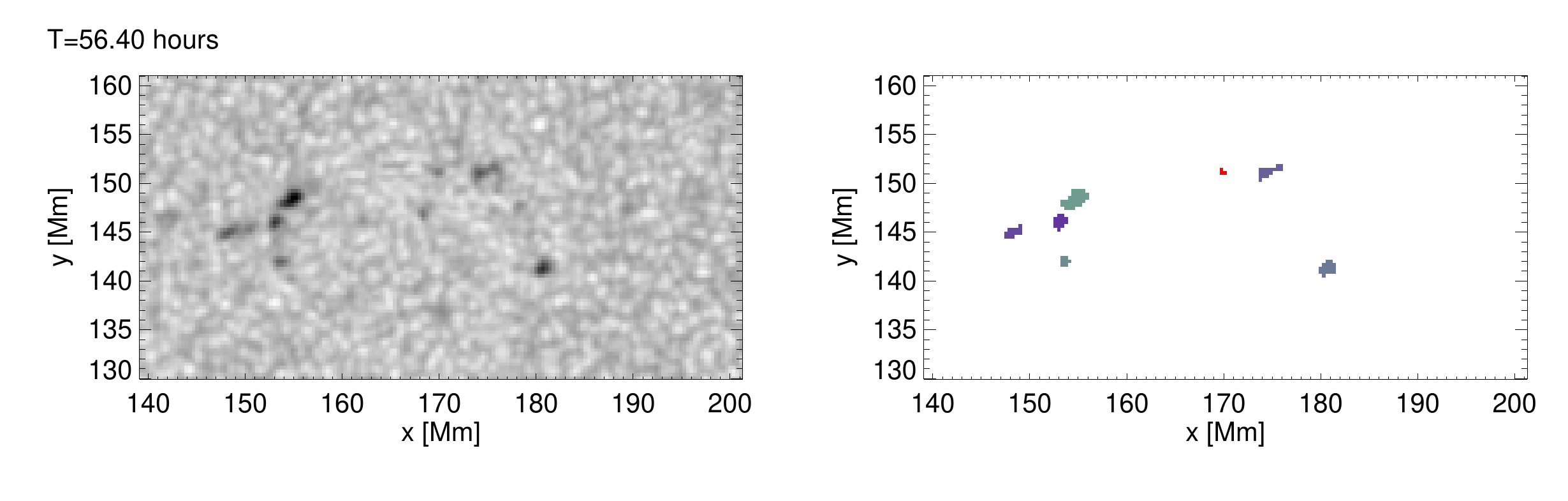}\\
    \includegraphics[width=0.8\textwidth]{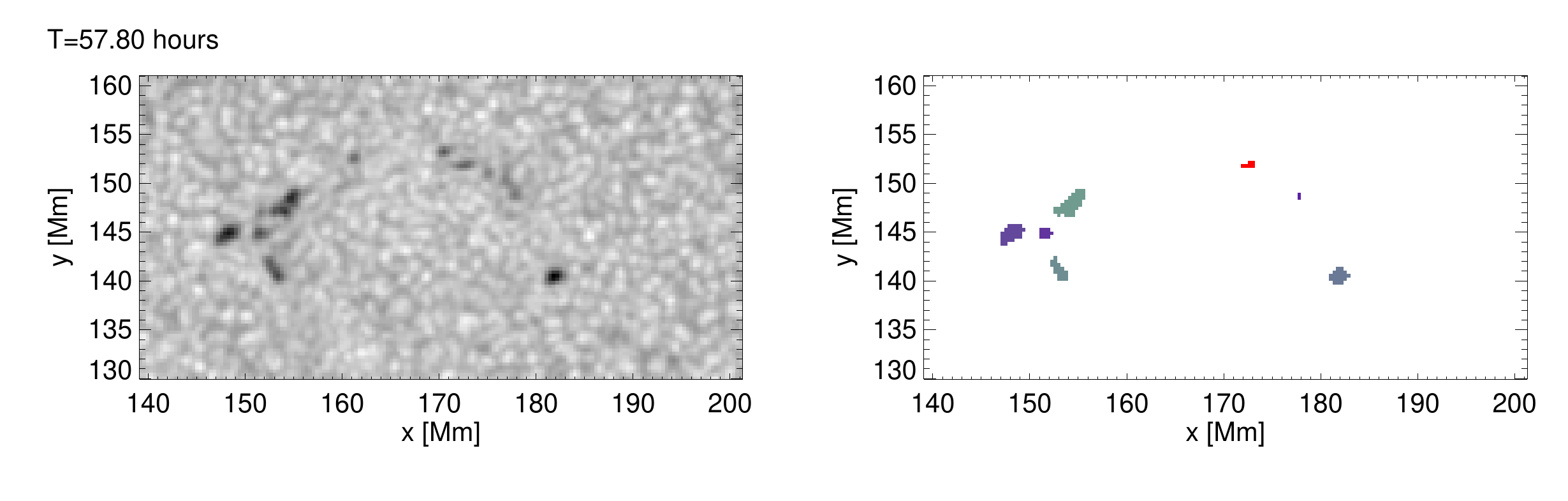}\\
    \includegraphics[width=0.8\textwidth]{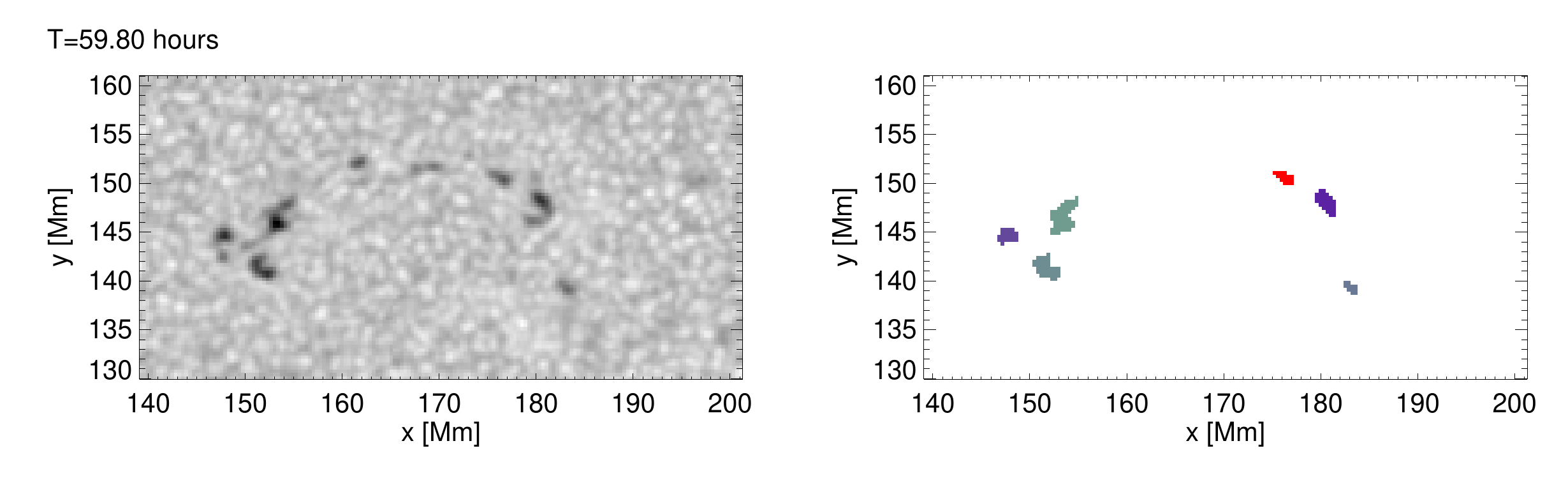}\\
    \includegraphics[width=0.8\textwidth]{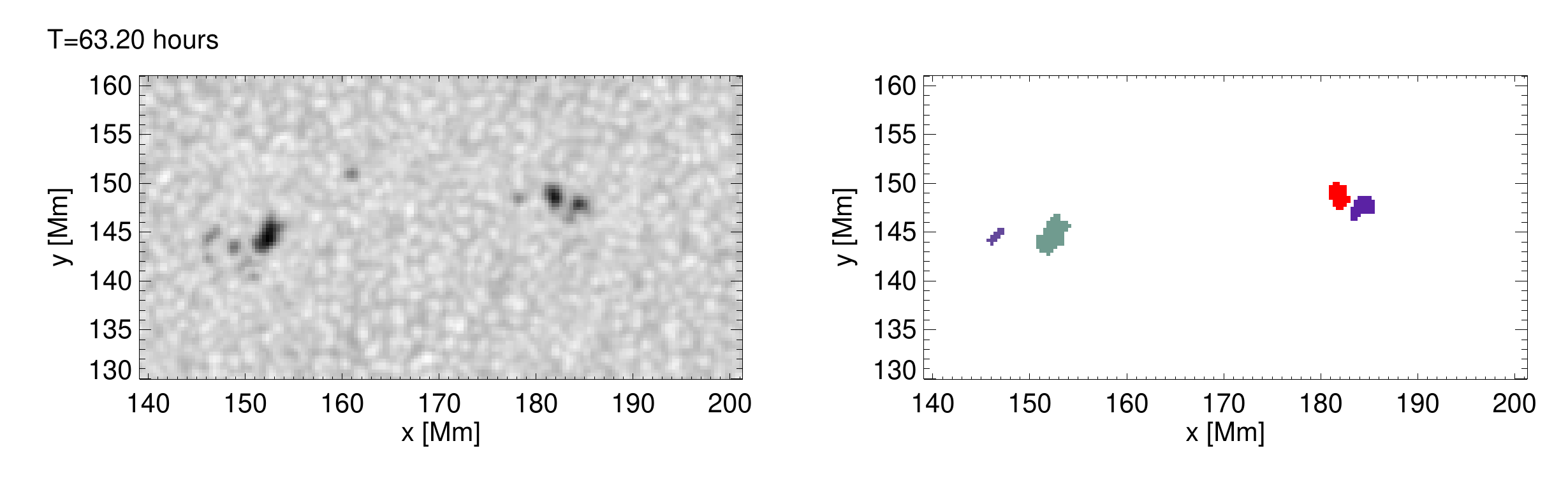}
    \caption{Example snapshots from fragment detection and tracking. In the left column, one can see the cut-out from the full field of view. In the right column, the detected fragments passing the selection criteria are shown; their identity is indicated by colours. None of the detected objects over the field of view were classified as a sunspot by the sunspot detection pipeline. There are many dark features in the pseudo-continuum images that are not marked as objects. These unmarked features did not meet the criteria for the lifetime or size.}
    \label{fig:NOAA11076_merge}
\end{figure*}

An example of object merging is given in Fig.~\ref{fig:NOAA11076_merge}. A demonstrable merging occurs around coordinates $x=180$~Mm and $y=150$~Mm. At $T=56.40$~hours, the first of the followed fragments (red) is born; at $T=57.60$~hours, the second one is born (purple). At $T=59.80$~hours, the fragments are stable, growing and travelling ahead of each other. The time $T=63.20$~hours is the last frame when the fragments were separated; in the next frame, they merge into one object, which kept the identity of the `red' fragment. We note that the blue-grey object on coordinates $x\sim 180$~Mm and $y\sim 140$~Mm, which emerged at $T=54.80$~hours (not displayed) grew in intensity between $T=56.40$~hours and $T=57.60$~hours, then it started to fade out and disappeared at $T=60.40$~hours (not displayed) without merging. 

The splitting event is considered if two objects in a given frame overlap at least one pixel each with a single object in the preceding frame. This means that a new object is born just next to a previously existing one. The label of the mother object is kept by the darker of the daughter objects and the brighter one is given a new label.

\begin{figure}
    \centering
    \includegraphics[width=0.5\textwidth]{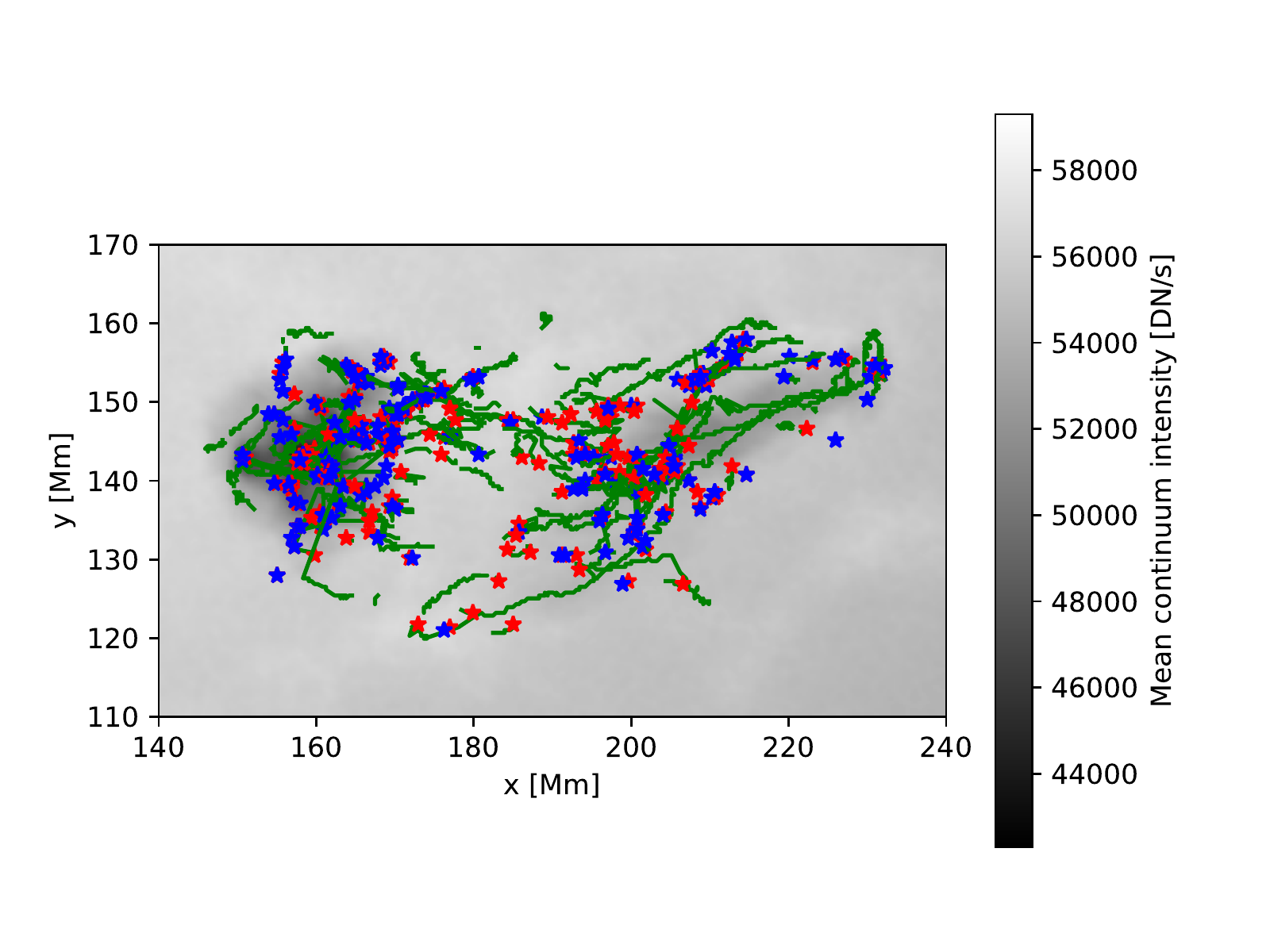}
    \caption{Fragments in AR NOAA 11076. The background image consists of an intensitygram averaged over the whole datacube time span. The whole considered field of view is plotted. The tracks of individual fragments (green lines) and the positions of the merging (red) and splitting (blue) events are overplotted. }
    \label{fig:NOAA11076_tracks}
\end{figure}

The merging and splitting events are recorded to a table, which contains information about the two objects considered, the time when the event occurred, and the type of the event. This allows one to create a spatio-temporal map of the merging and splitting centres. One may also draw a historic trajectory of each object from its birth to its decay or merging. An example is given in Fig.~\ref{fig:NOAA11076_tracks}. Here one can see that most of the features are detected within the dark, that is, magnetised regions. 

In principle, the merging and splitting events may occur anywhere in the active region. However, if the scenario by \citeauthor{1987SoPh..112...49G} is valid, one would anticipate that merging and splitting events occur around preferred locations, likely near the locations of sunspots. To investigate this issue, we applied a cluster analysis to the locations of the merging and splitting events. 

We utilised the density-based spatial clustering of applications with noise \citep[DBSCAN,][]{ester1996densitybased} implemented within Python's Scikit-Learn library. DBSCAN is a density-based non-parametric clustering algorithm. Given a set of points in some space, it groups together points that are closely packed together (points with many nearby neighbours). Points lying alone in low-density regions (with nearest neighbours far away) are marked as noise. The advantage of the DBSCAN algorithm is that the expected number of clusters does not have to be known a priori. The algorithm can also find arbitrarily shaped clusters.

\begin{figure}
    \centering
    \includegraphics[width=0.5\textwidth]{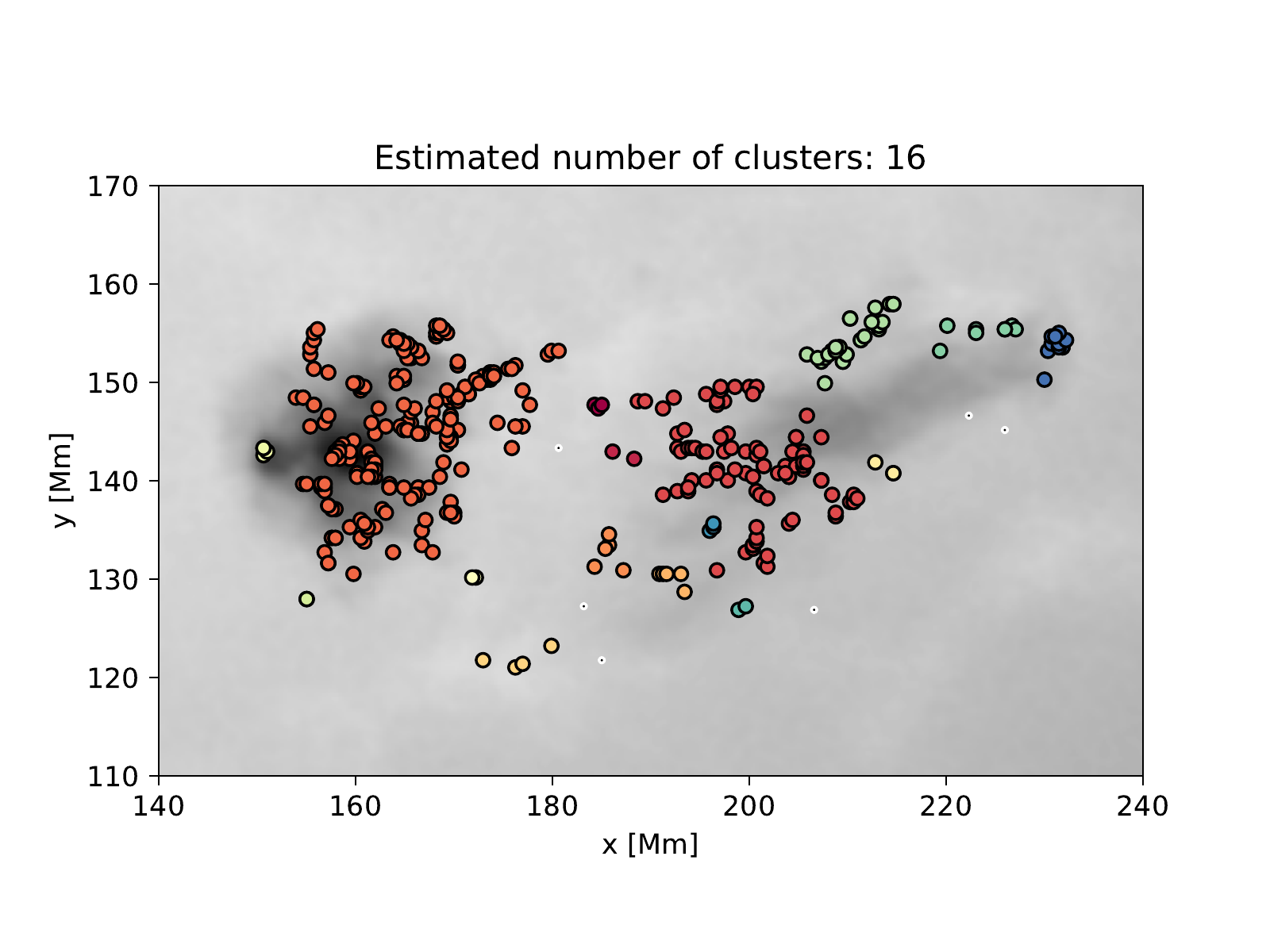}
    \caption{Clustering of merging and splitting events over the portion of the field of view of NOAA 11076. Different colours indicate detected clusters. Black points with white strokes represent points which do not belong to any of detected clusters. }
    \label{fig:NOAA11076_clusters}
\end{figure}

\begin{figure*}
    \centering
    \includegraphics[width=\textwidth]{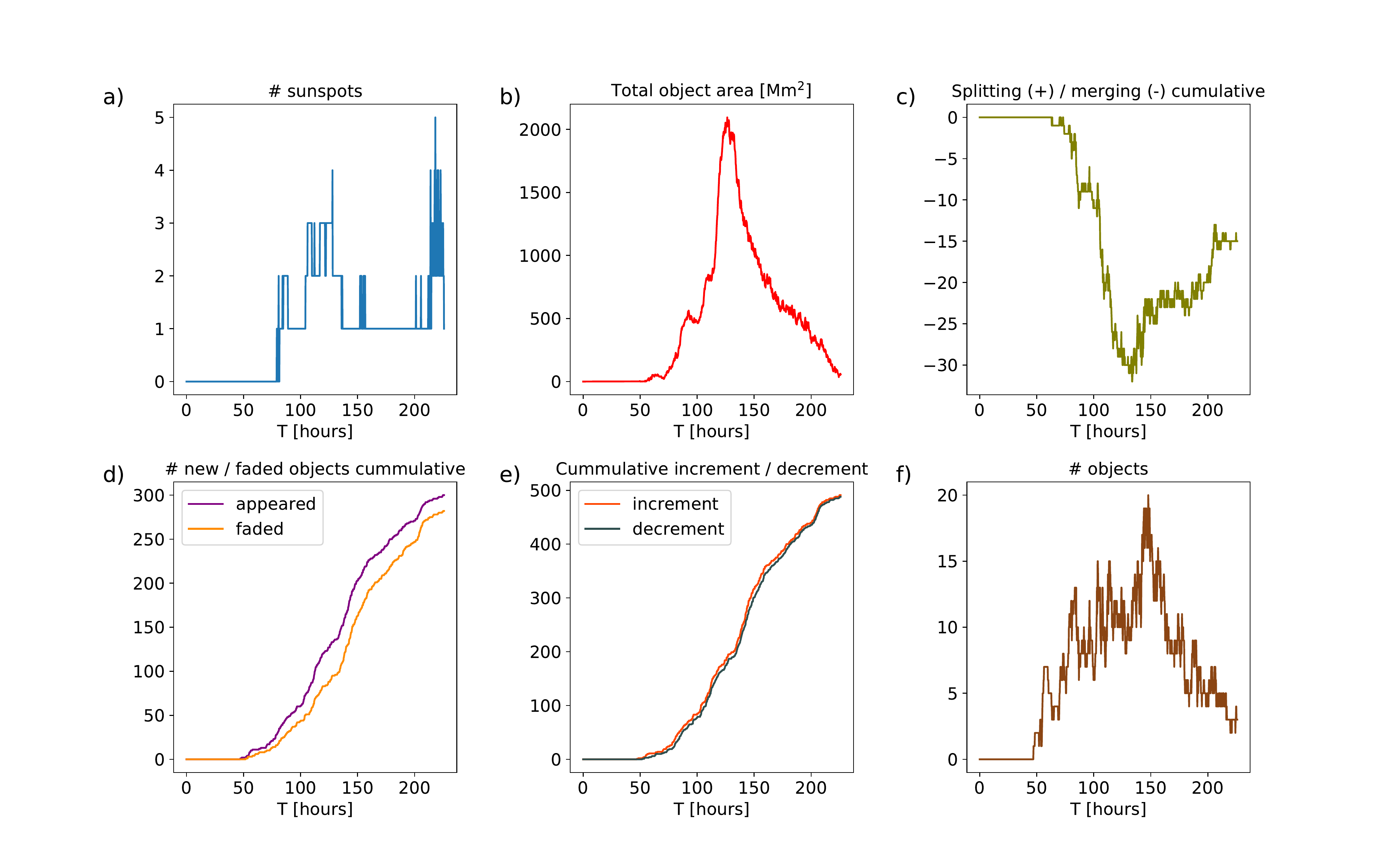}
    
    \caption{Time evolution of various quantities describing the sunspots and fragments in NOAA~11076. One can see the evolution of the number of detected sunspots (a) and their area (b). The cumulative curve of the splitting or merging events is also plotted (c), showing which of the quantities prevailed until the given moment. In panel d), one can see the cumulative curves of the number of objects that newly appeared and that faded staying alone. In panel e), one can see the cumulative balance between the increase and decrease in the number of objects. Finally, in panel f), the total number of the detected objects at the given time is plotted. }
    \label{fig:NOAA11076_lightcurves}
\end{figure*}

For each datacube, we identified the clustering of the merging and splitting events. We used the minimum distance of the cluster members $\varepsilon=10$~px and the minimum number of cluster members of 2. As a distance measure, we used the 2D Euclidian norm. We did not consider a factor of time in the distance metric; therefore, the splitting and merging events belonging to the same identified cluster did not have to occur at the same time. 

An example of the identified clustering is given in Fig.~\ref{fig:NOAA11076_clusters}. This is a typical example where one can see that most of the splitting and merging events are located within the regions with sunspots. 

\section{Results}

\subsection{Characteristics of active regions}
For each active region, we obtained several data products as functions of time. Firstly, the number of detected individual sunspots, in case there are multiple umbrae within one penumbra, they count as one sunspot. Secondly, we measured the total area taken by the detected objects. Thirdly, we put together an overview of the fragments in each frame including (a) newly appearing fragments, (b) fragments resulting from the splitting of the mother fragment, (c) fragments disappearing because they merged with other fragments, and (c) fragments that faded staying alone. These quantities were evaluated not only by the frame but also in their cumulative forms, that is the sum of the quantity from the beginning of the sequence to the given timestamp matching the stamp of the given frame. 

\begin{figure*}
    \centering
    \includegraphics[width=0.49\textwidth]{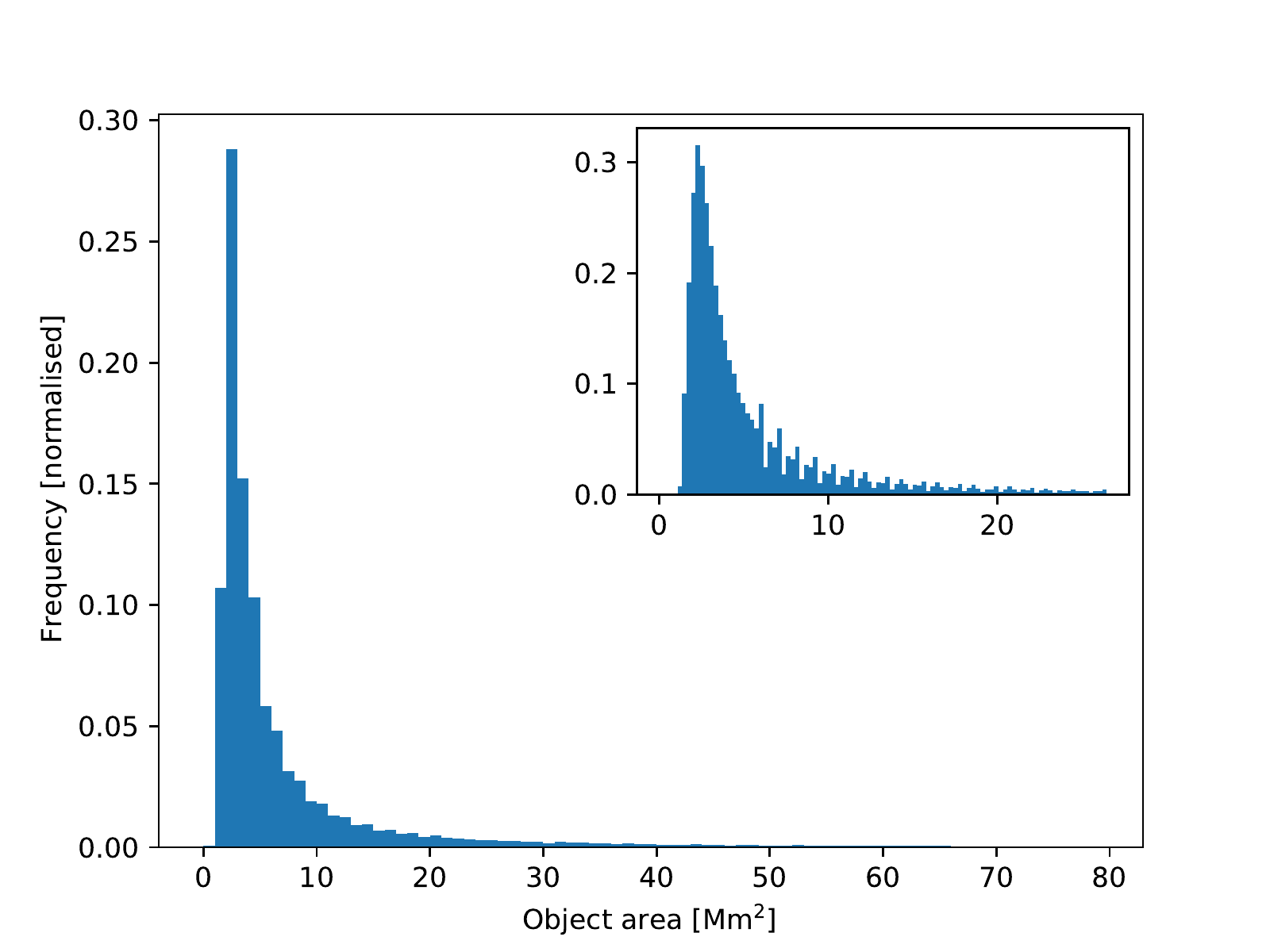}
    \includegraphics[width=0.49\textwidth]{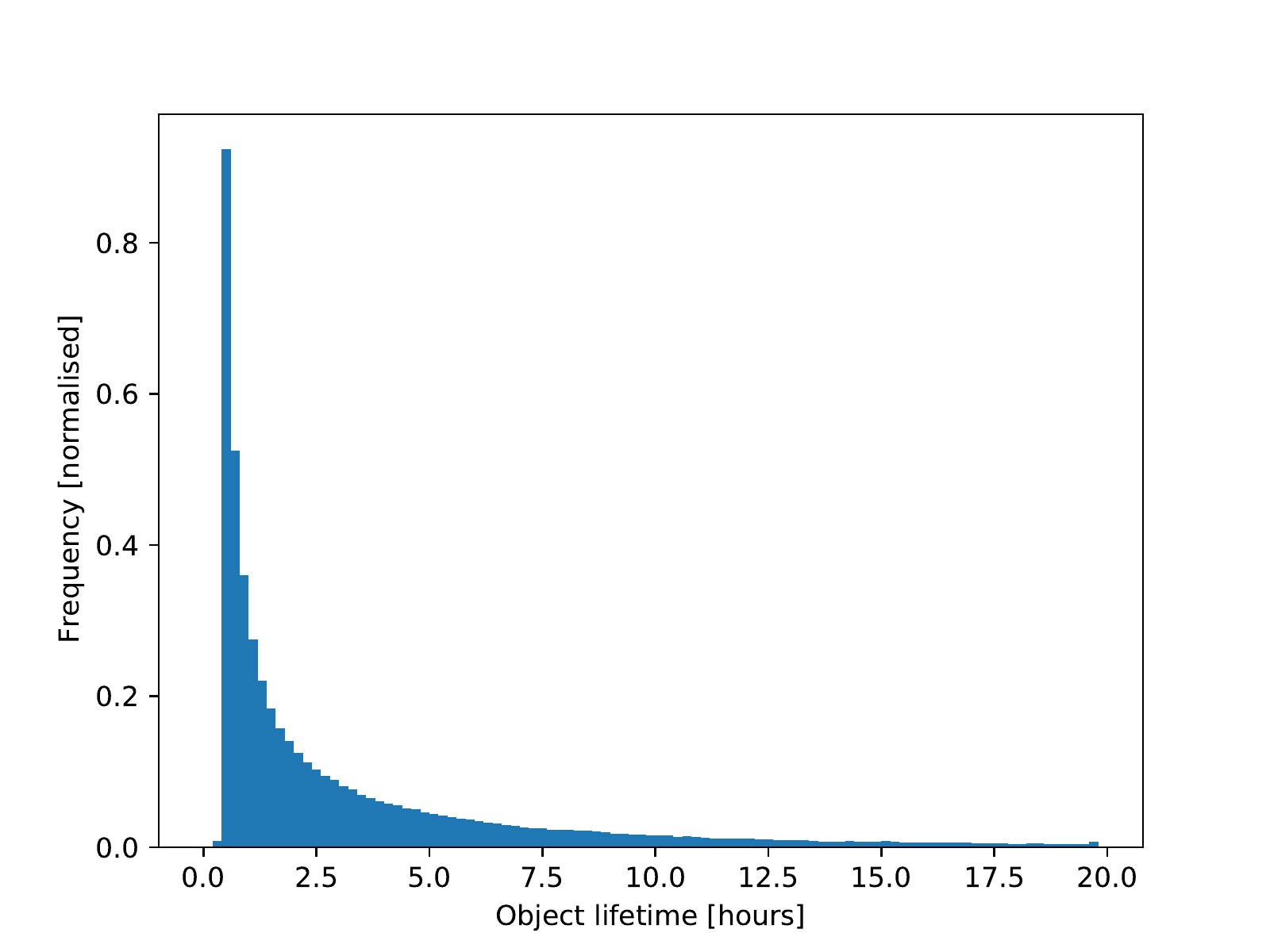}
    \caption{Left: Histogram of the object sizes (in terms of the areas) computed for the binning of 1~Mm$^2$. The inset shows the magnified histogram for small sizes with the bin size of 0.267~Mm$^2$, which is twice the natural sampling. Right: Histogram of the object lifetimes. }
    \label{fig:histograms_objects}
\end{figure*}

An example for one active region NOAA 11076, where we captured not only its formation but also a significant part of its decay, is shown in Fig.~\ref{fig:NOAA11076_lightcurves}. The region reached its maximum area at a time of $T\sim 120$~hours (red line, top middle panel). Before this time, merging events (negative) were dominant over the active region (olive line, top right panel); after this moment, the splitting (positive) events prevailed. The total number of detected objects in the field of view reached a local minimum at $T\sim 120$~hours, whereas the maximum number of objects was counted about 30~hours later. The slope of the olive curve indicates that between these two instances, the fragmentation of the objects was particularly fast. 

Interestingly, the total number of merging events did not equal the total number of splitting events since the olive line in Fig.~\ref{fig:NOAA11076_lightcurves} remained negative. The remaining objects faded without further fragmentation. On the other hand, at each moment, the total number of appearing objects (new objects and objects resulting from splitting events) and disappearing objects (faded objects and mergers) were in a very close balance (bottom middle panel of Fig.~\ref{fig:NOAA11076_lightcurves}). This would indicate a relatively short expected lifetime of the objects. The object appears and soon after merges or fades. 

\subsection{Fragment statistics}
In total, the object tracking pipeline identified 303\,918 objects. The histograms of their areas and lifetimes are given in Fig.~\ref{fig:histograms_objects}.

It would seem that small objects clearly prevail, which also usually have a short lifetime. There is a peak in the histograms of areas of the detected objects around an area of 2~Mm$^2$, which could be considered as the typical object size. This value corresponds to roughly 15~px$^2$, thus this is not noise. The inset histogram with the bin size of 0.267~Mm$^2$, which is twice as large as the data sampling, shows the details around the peak. The lifetime, on the other hand, seems to continuously decrease from about 24 minutes (2 frames at a cadence of 12~min per frame) to larger values. 

\begin{figure}
    \centering
    \includegraphics[width=0.49\textwidth]{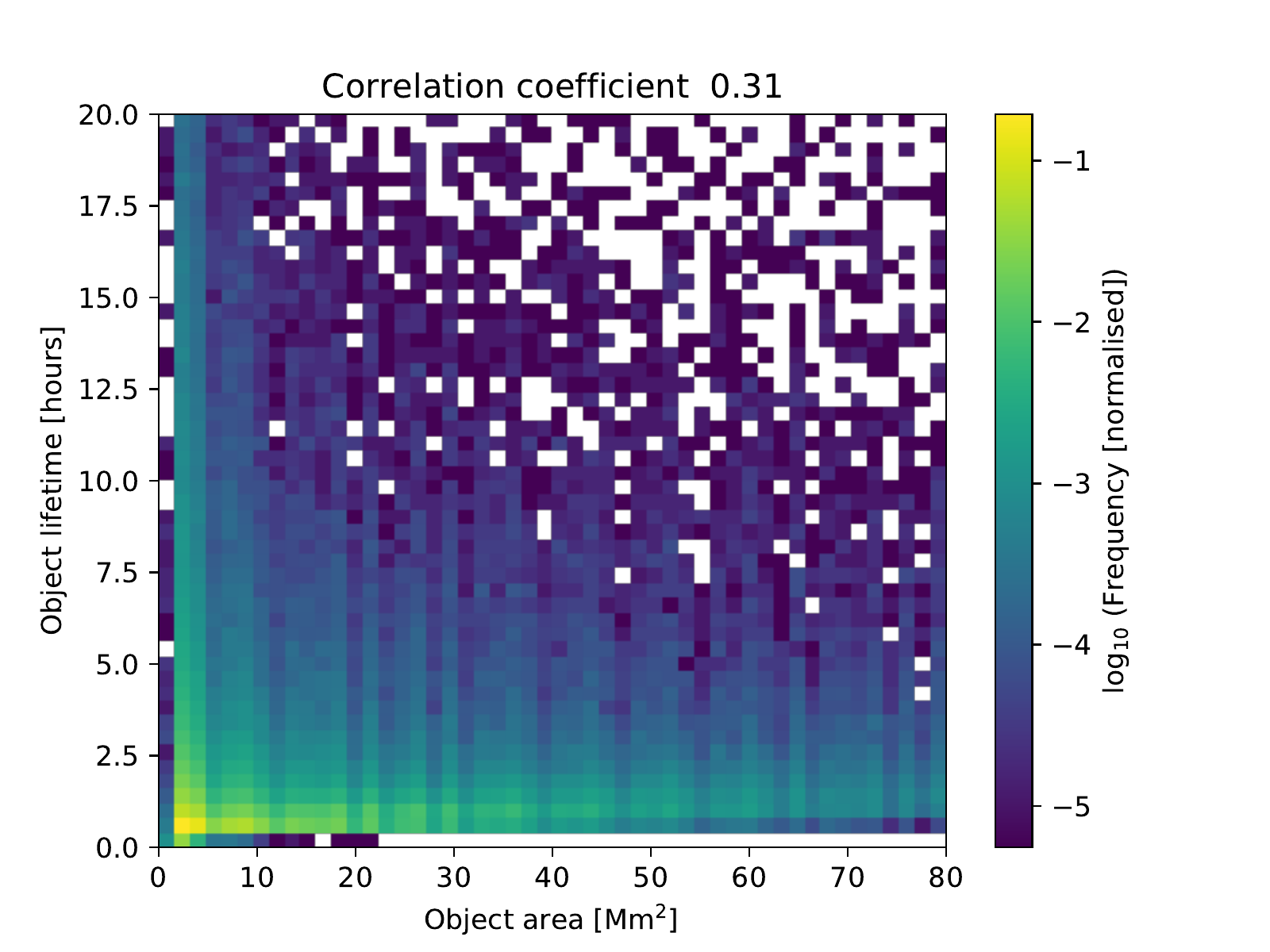}
    \caption{2D histogram of object area and lifetime. The histogram is plotted in the logarithmic scale. It indicates that there is no significant correlation between the size and the lifetime of the objects. }
    \label{fig:histograms_objects_2D}
\end{figure}

The area and the lifetime of the objects only correlate weakly with a correlation coefficient of about 0.3. The density histogram of this dependence is given in Fig.~\ref{fig:histograms_objects_2D}. We note that the colour scale is plotted in a logarithmic scale; otherwise, the plot would be strongly dominated by a peak in the lower-left corner. 

The parameters of the individual fragments are strongly affected by our methodology, which is especially true for their lifetimes. During the splitting event, there are more choices as to how to select the successor of the mother object. Our choice of the darker (there we can assume a stronger magnetic field) fragment being the successor performs well, except for the type of event when a small dark object splits from an evolved sunspot. An area averaged intensity of the sunspot with a penumbra might be larger (due to the relative bright large-area penumbra) than the area-averaged intensity of the small fragment, hence the small fragment keeps the label of the mother object. In the extreme case of the sunspot with an extended penumbra, it may change the identity due to splitting frame by frame. Fortunately, in our sample, this does not occur too often overall. After removing objects with a lifetime shorter than 1~hour and a size larger than 706 Mm$^2$ (this roughly corresponds to a size of a supergranule), the correlation coefficient between the area and the lifetime is about 0.5. The consecutive analysis is not affected by this bias in the lifetime of the objects. 

\subsection{Location of merging and splitting events with respect to sunspots}
\label{sect:locations}
\begin{figure*}
    \centering
    \makebox[0.46\textwidth]{Forming ARs} \makebox[0.46\textwidth]{Decaying ARs}\\
    \raisebox{5.5cm}{a)}\includegraphics[width=0.46\textwidth]{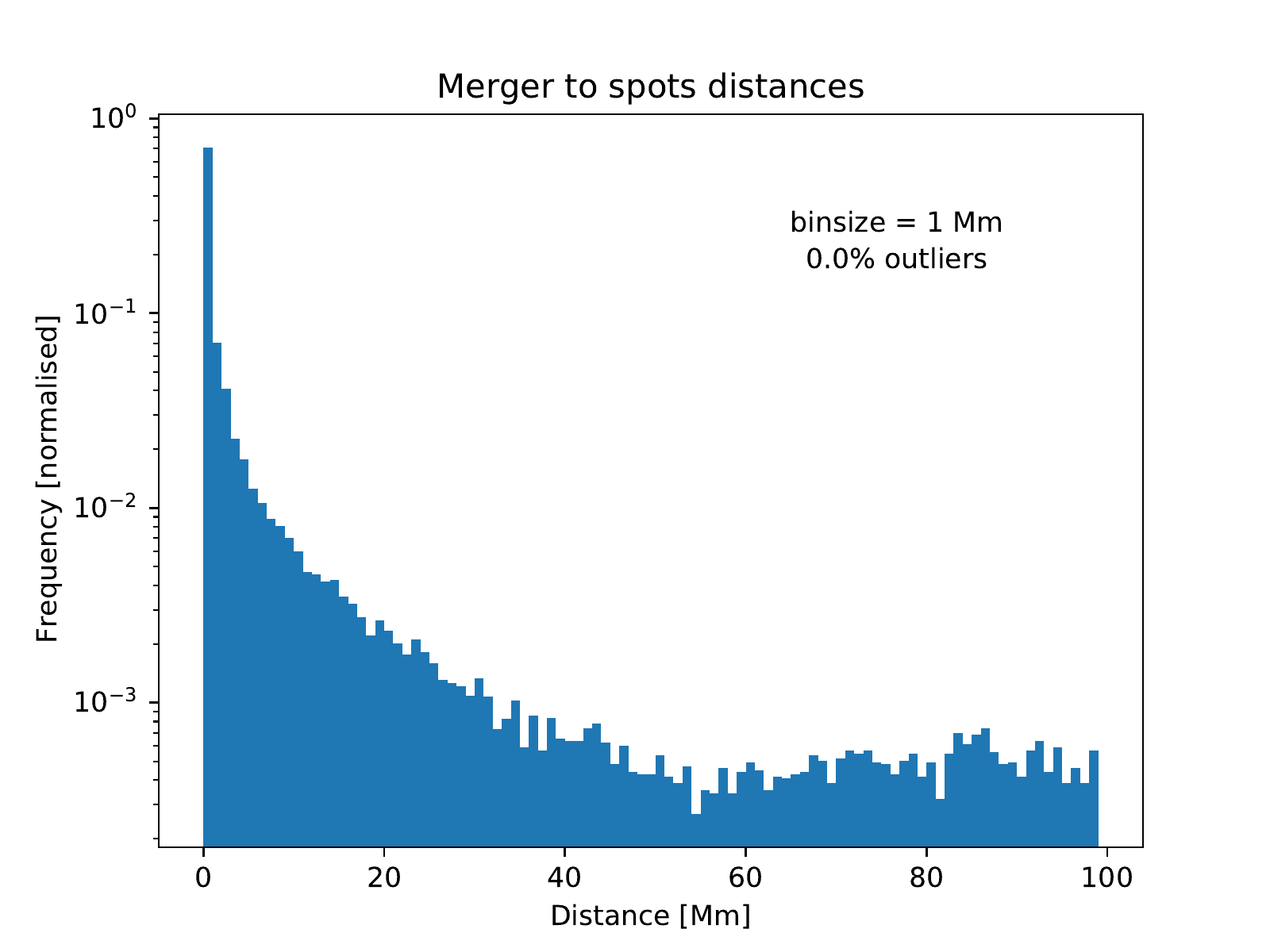}
    \raisebox{5.5cm}{b)}\includegraphics[width=0.46\textwidth]{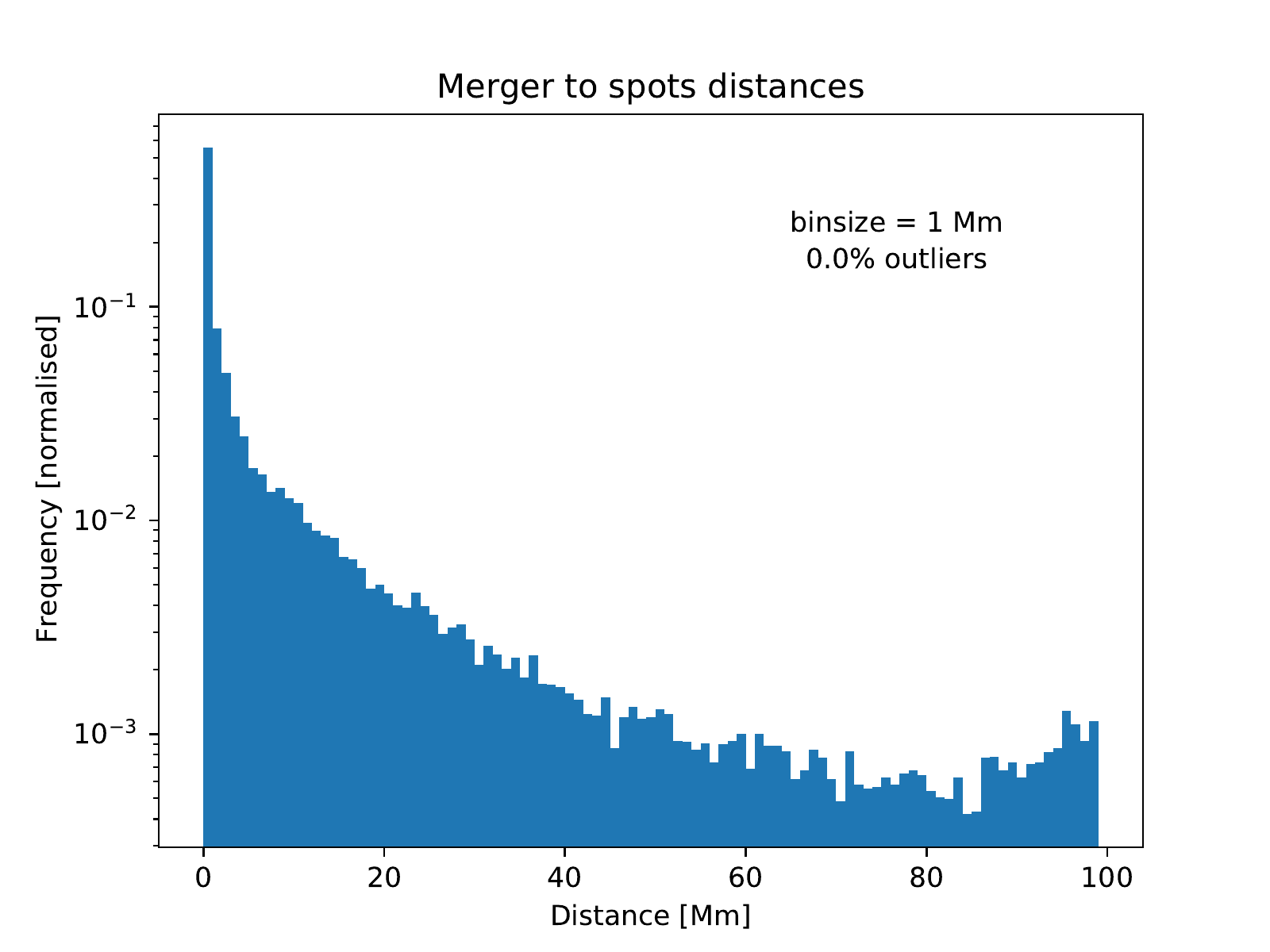}\\    
    \raisebox{5.5cm}{c)}\includegraphics[width=0.46\textwidth]{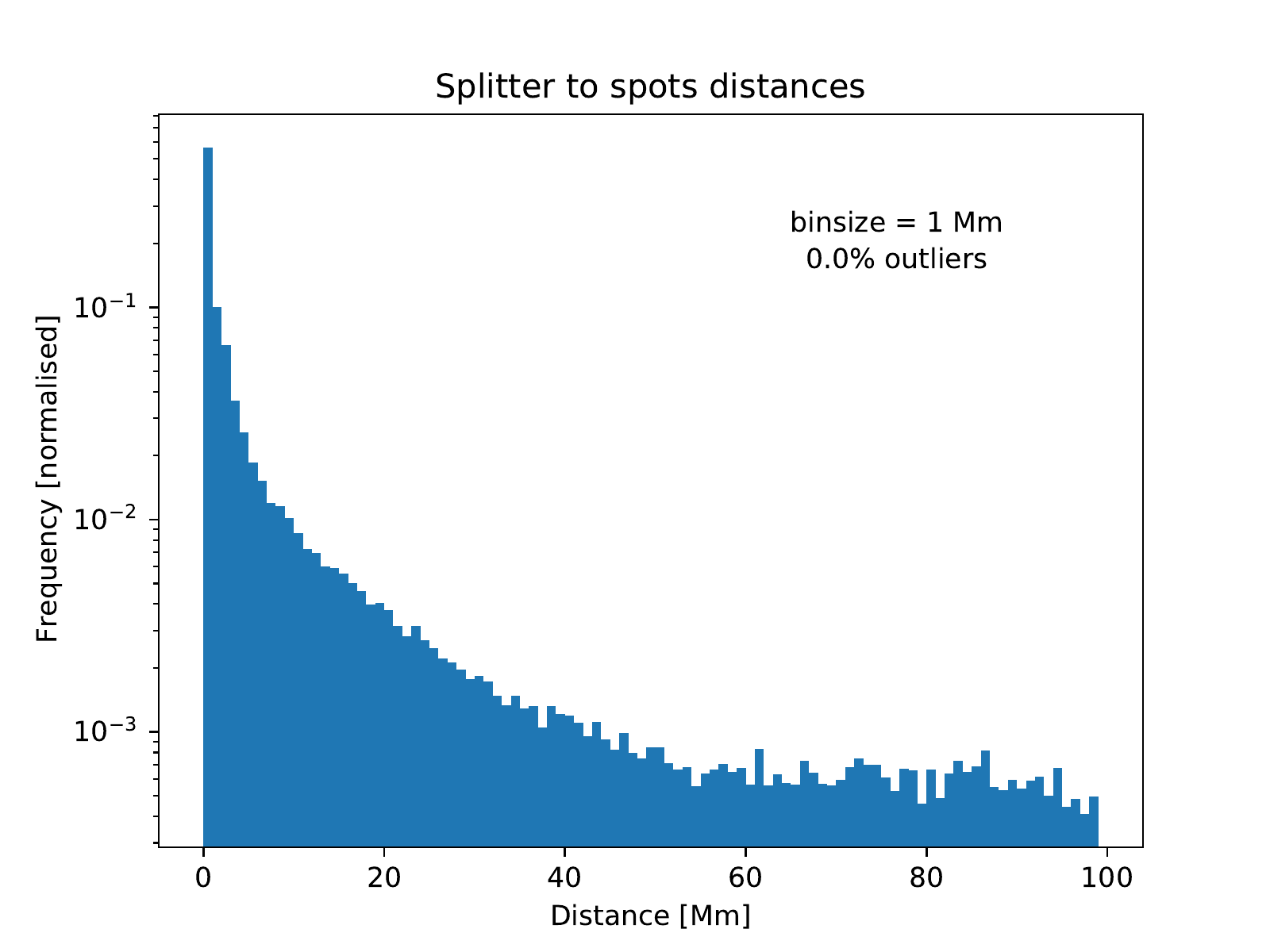}
    \raisebox{5.5cm}{d)}\includegraphics[width=0.46\textwidth]{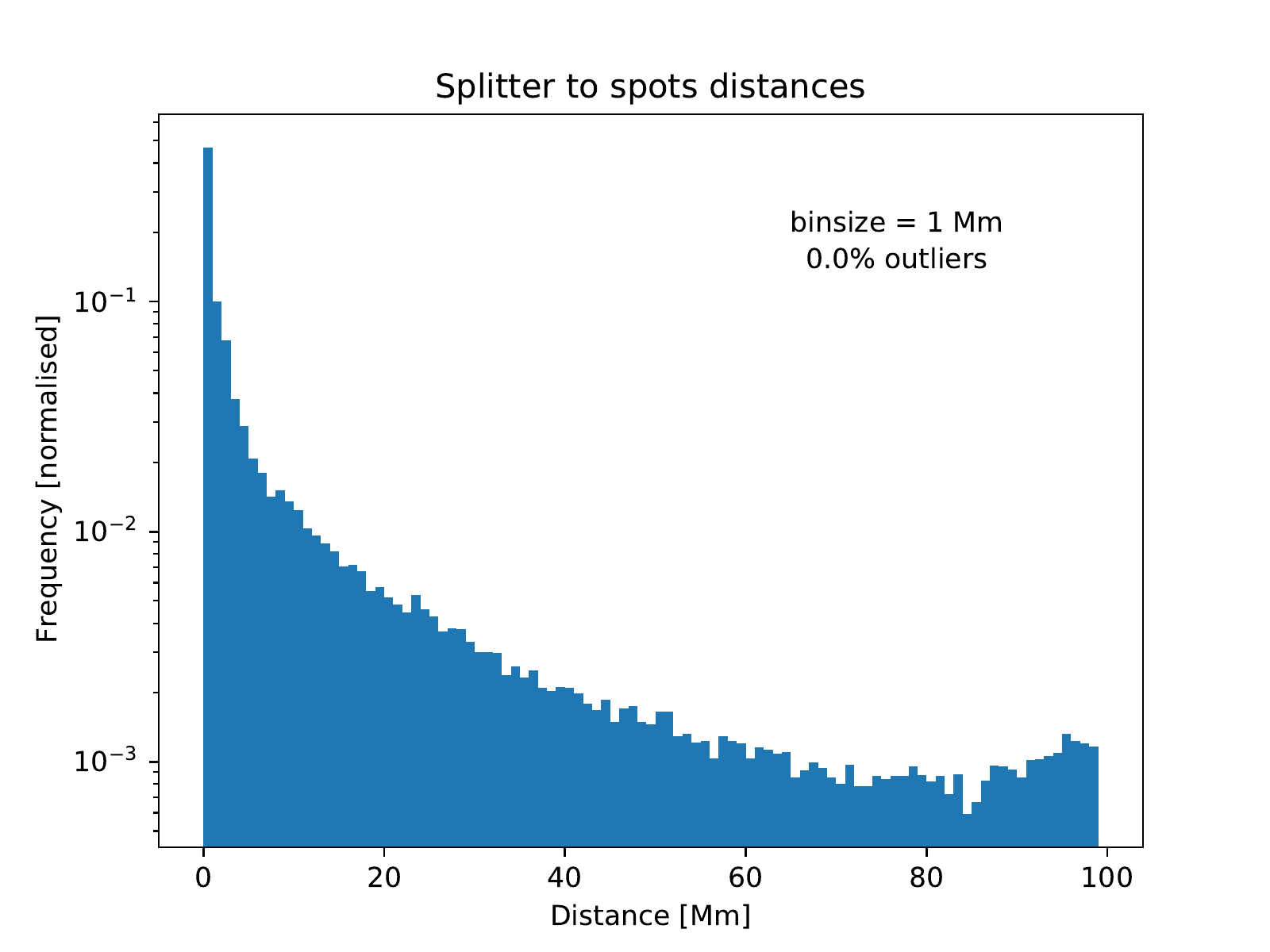}\\    
    \raisebox{5.5cm}{e)}\includegraphics[width=0.46\textwidth]{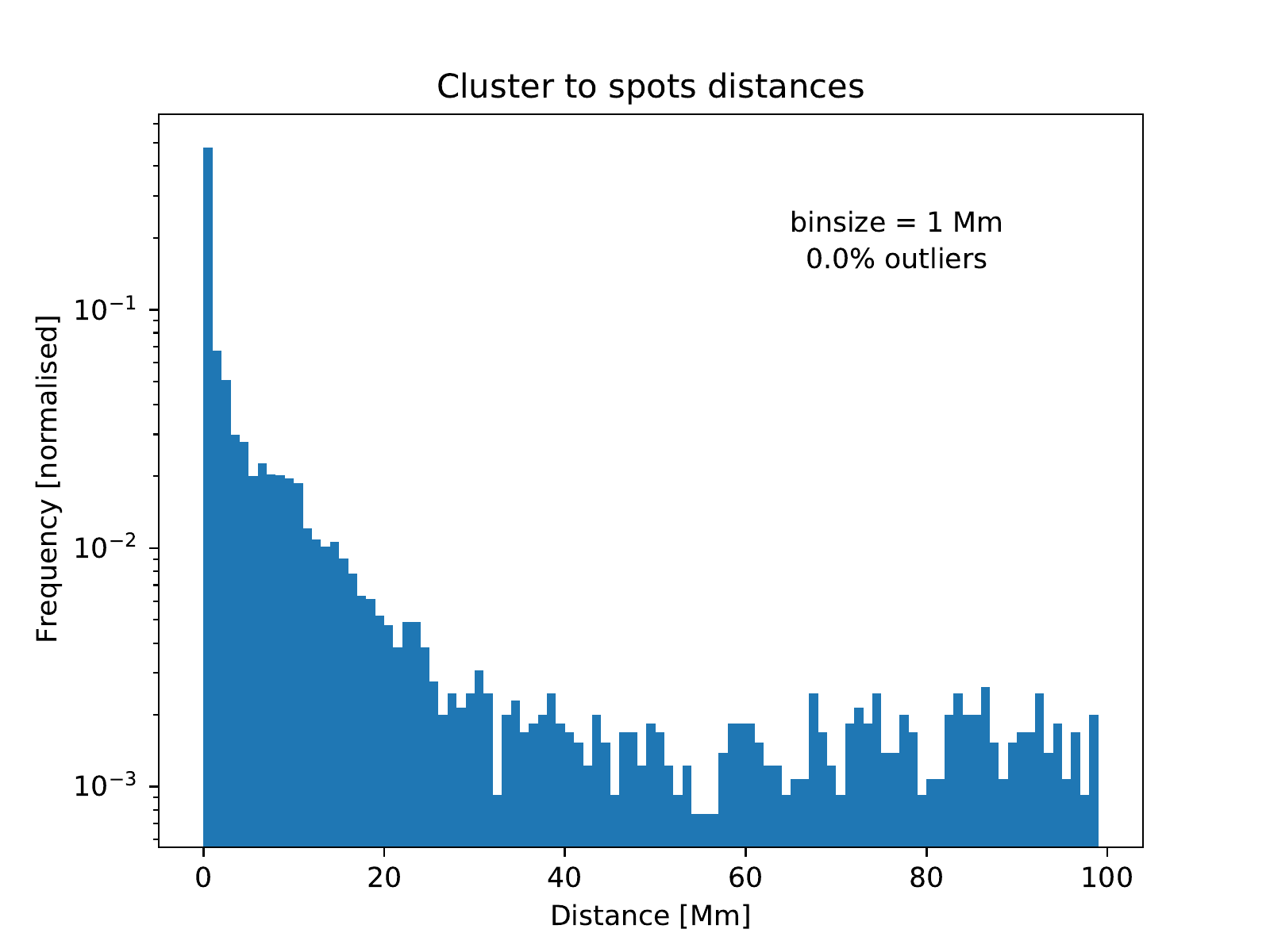}
    \raisebox{5.5cm}{f)}\includegraphics[width=0.46\textwidth]{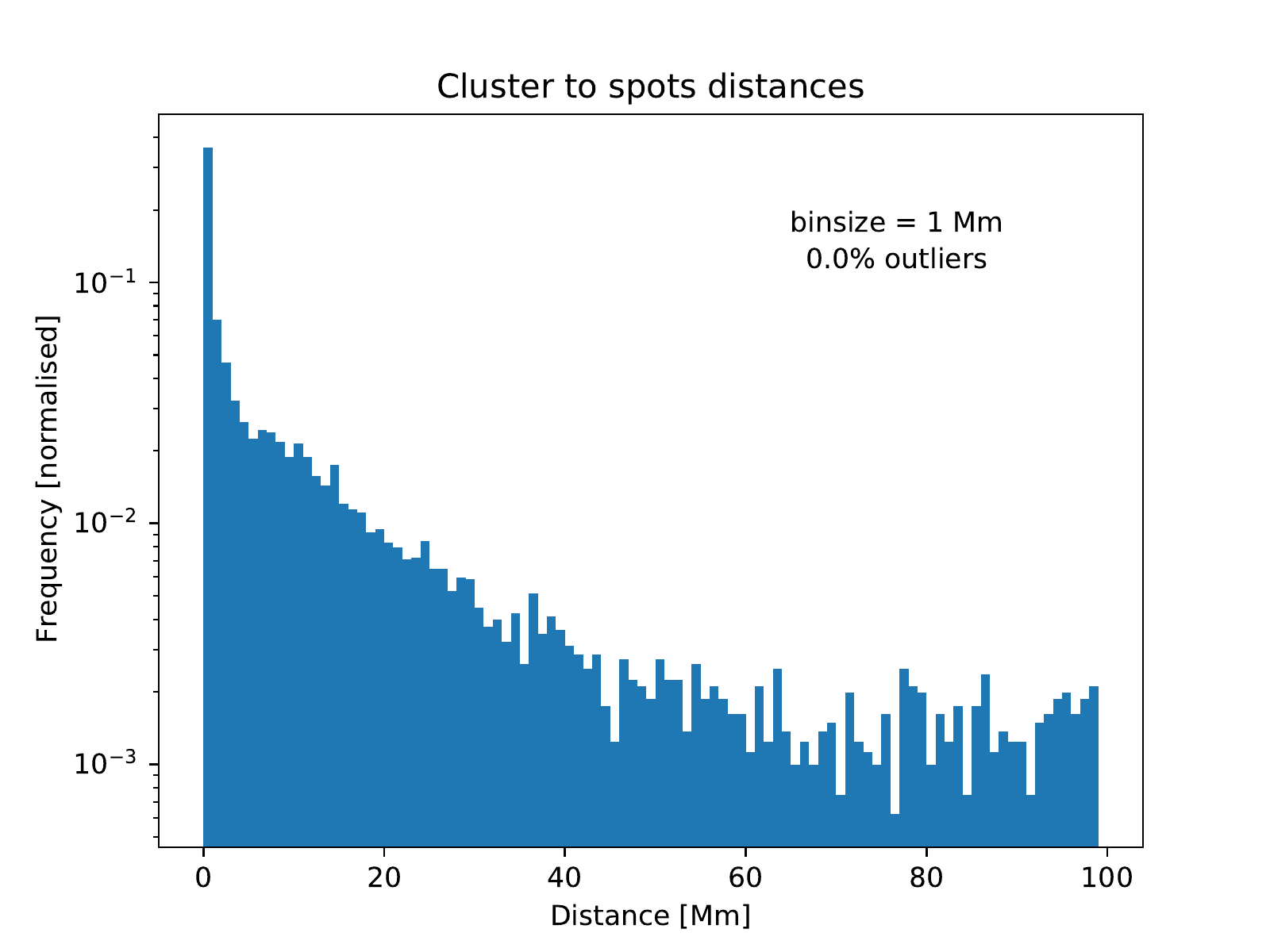}\\
    \caption{Histograms of distances between the positions of the merging events to the closest sunspot (top), position of the splitting events to the closest sunspot (middle), and between the gravity centres of the merging and splitting clusters and the closest sunspot (bottom). The histograms are plotted for the active regions in their emerging and formation stage (left) and for the decaying active regions (right). It is important to take note of the logarithmic scales on the vertical axes. }
    \label{fig:histograms_distances}
\end{figure*}

According to the scenario by \cite{1987SoPh..112...49G}, the sunspot is formed by merging events from the fragments. That would indicate that the distance between the merging events and the real sunspots should be rather small. Thus we investigated the histogram of distances between the positions of the merging events and the closest sunspot separately for forming and decaying active regions. This is displayed in Fig.~\ref{fig:histograms_distances}a and Fig.~\ref{fig:histograms_distances}b. Obviously, the merging events occur most
often next to the closest sunspot, thus supporting the scenario by \cite{1987SoPh..112...49G}. There seems to be a different distribution function for this quantity for the forming and decaying regions, where the width of the distribution in distances seems to be larger in the case of the decaying regions. This indicates that there are random interactions between fragments occurring away from the sunspots during the active regions' decay. 

The same holds true for the location of the splitting events (see Fig.~\ref{fig:histograms_distances}c,d). This would indicate that there is a slight preference of the merging and splitting events to occur very close to the sunspot location in the case of the forming active region, whereas when the active region is in the decaying phase, the connection is somewhat loose. Local helioseismology showed that there are large-scale inflows around active regions that are mostly present during the later emerging and stationary phases and they seem to weaken during the decaying phase \citep{2001ESASP.464..213H,2007ApJ...667..571K,2008ApJ...680L.161S}. The existence of these inflows during the formation phases and their lack during the decaying phase would explain our observation. Also, a wider distribution of distances counts the possibility of the emergence of the secondary polarities, which usually does not happen in the early phases of the active region evolution. 

Only with the aim of complementing the picture, we plotted histograms of distances for the gravity centres of the clusters of merging and splitting events from the closest sunspot in Fig.~\ref{fig:histograms_distances}e,f. Again, there is a clear preference for the correspondence of the positions of the event clusters with the location of sunspots. Not surprisingly, the connection is again somewhat loose in the case of the decaying active regions. 

\begin{figure*}
    \centering
    \makebox[0.46\textwidth]{Forming ARs} \makebox[0.46\textwidth]{Decaying ARs}\\
    \raisebox{5.5cm}{a)}\includegraphics[width=0.46\textwidth]{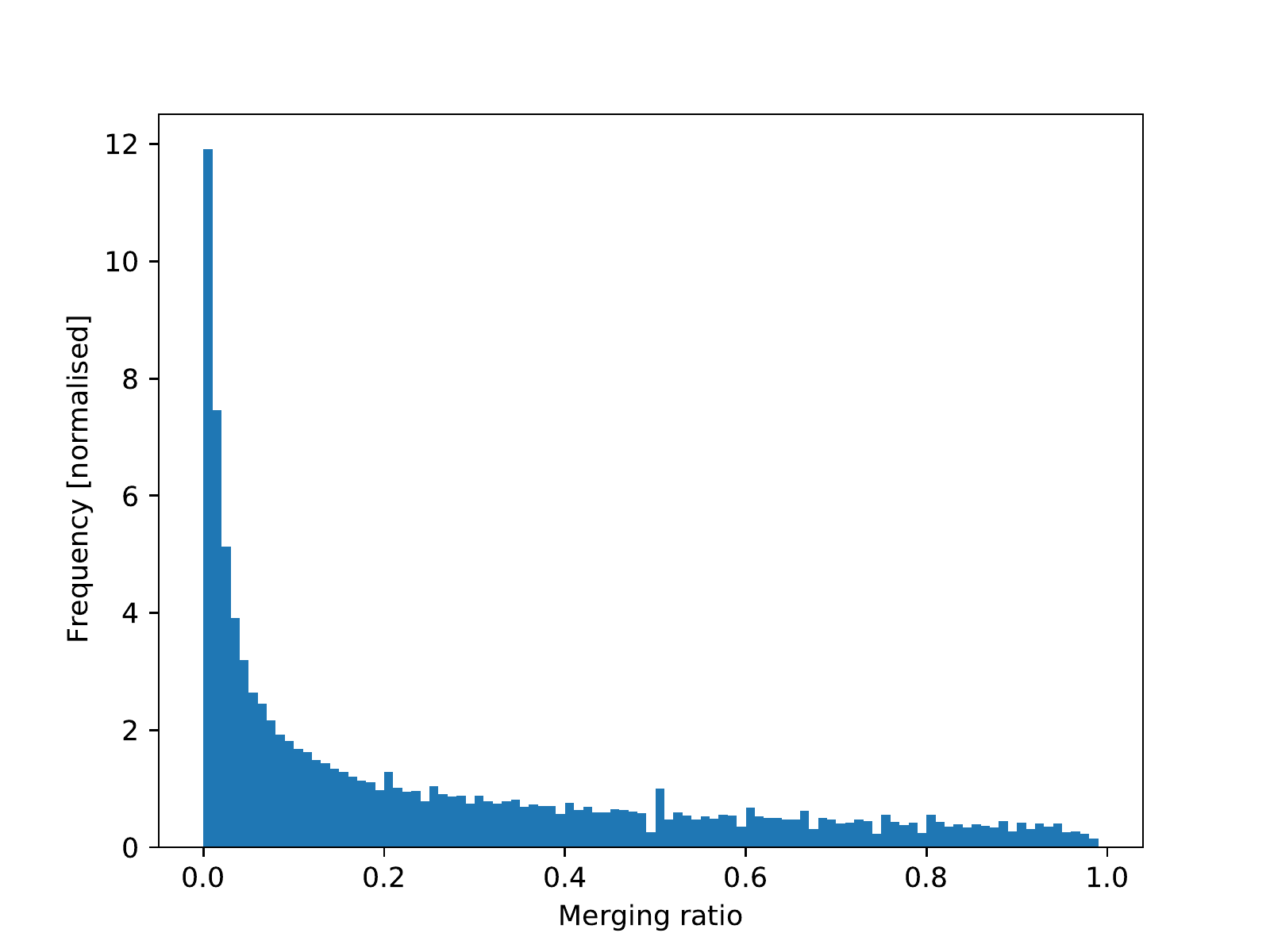}
    \raisebox{5.5cm}{b)}\includegraphics[width=0.46\textwidth]{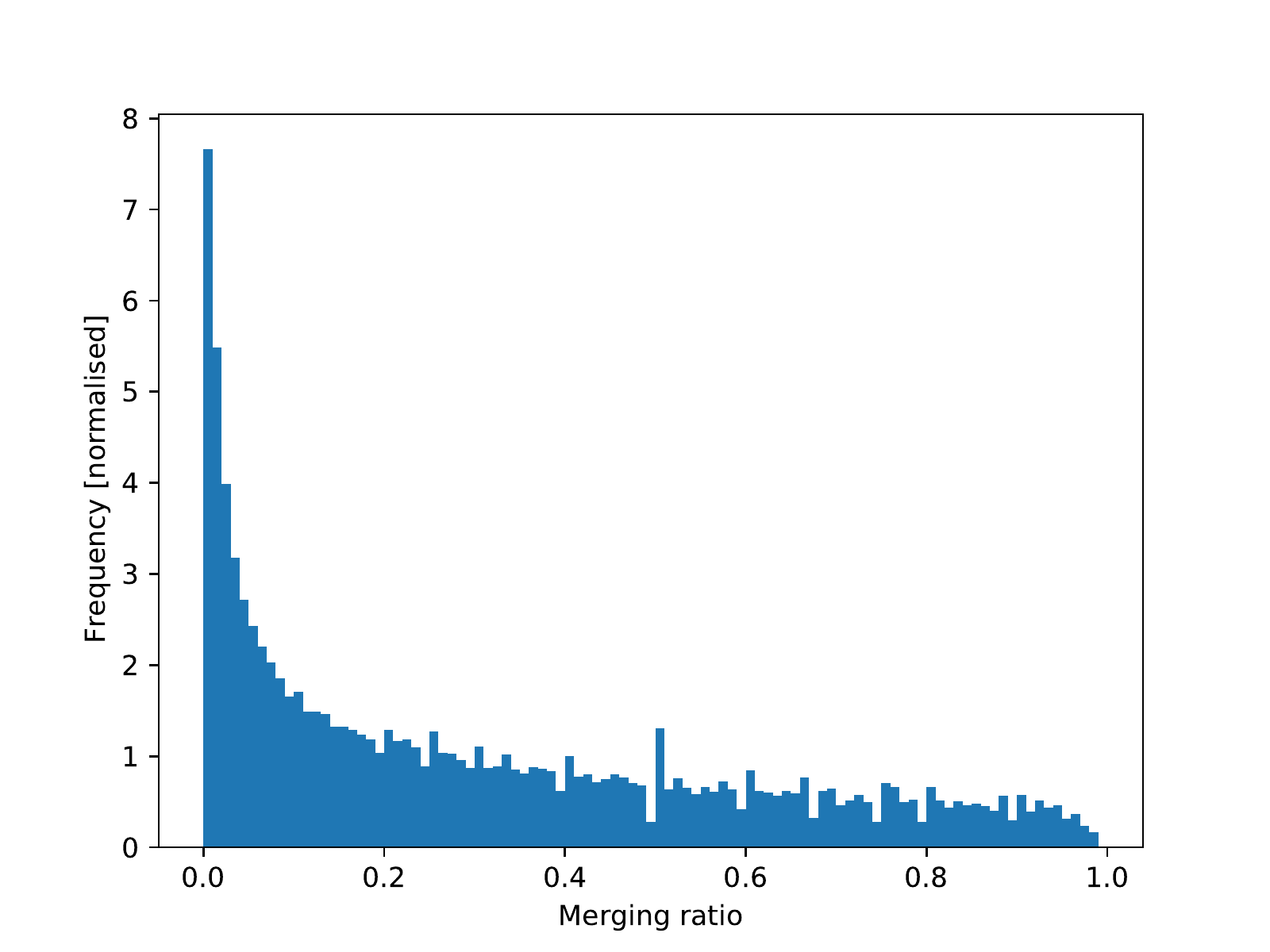}\\    
    \raisebox{5.5cm}{c)}\includegraphics[width=0.46\textwidth]{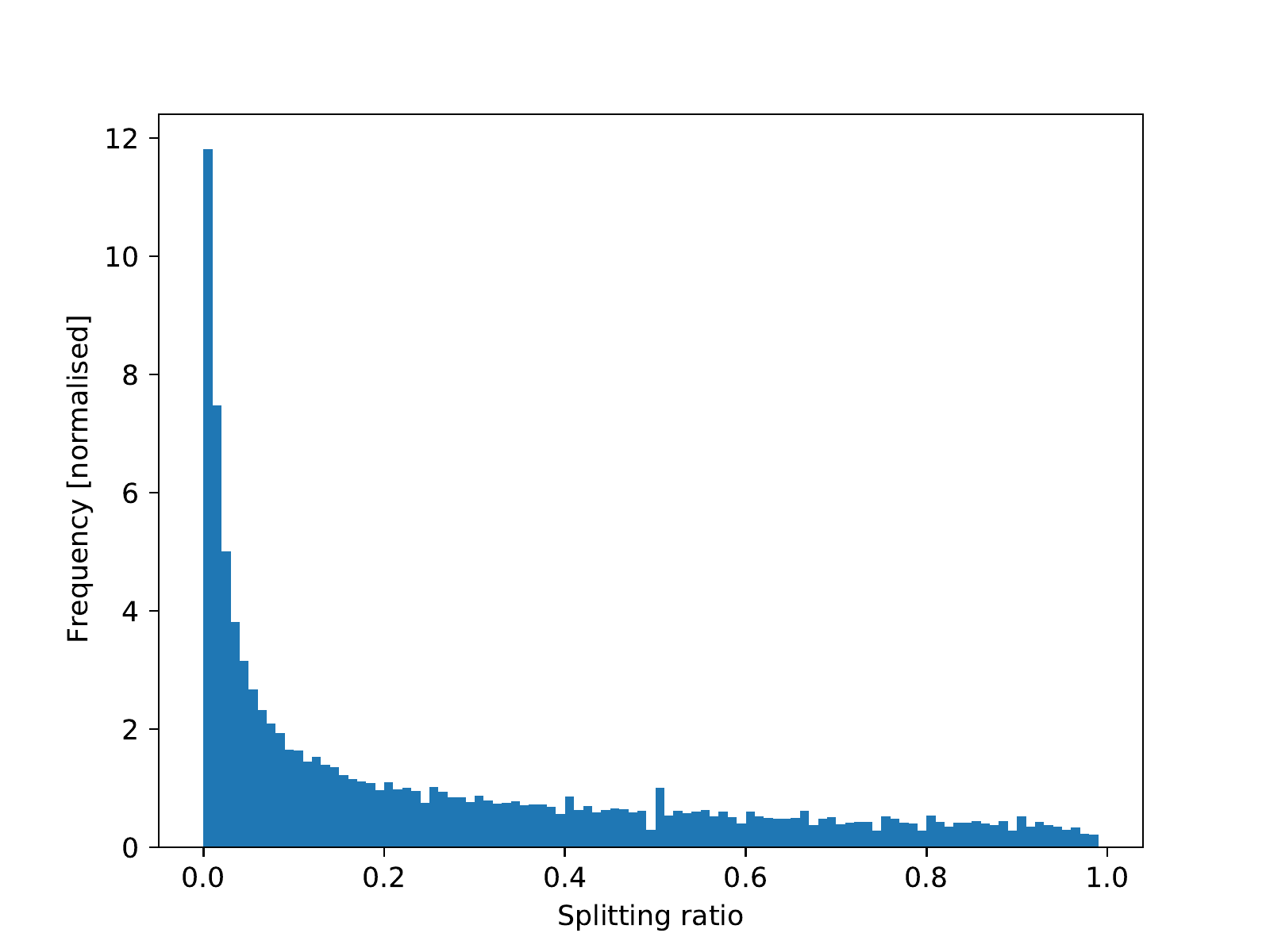}
    \raisebox{5.5cm}{d)}\includegraphics[width=0.46\textwidth]{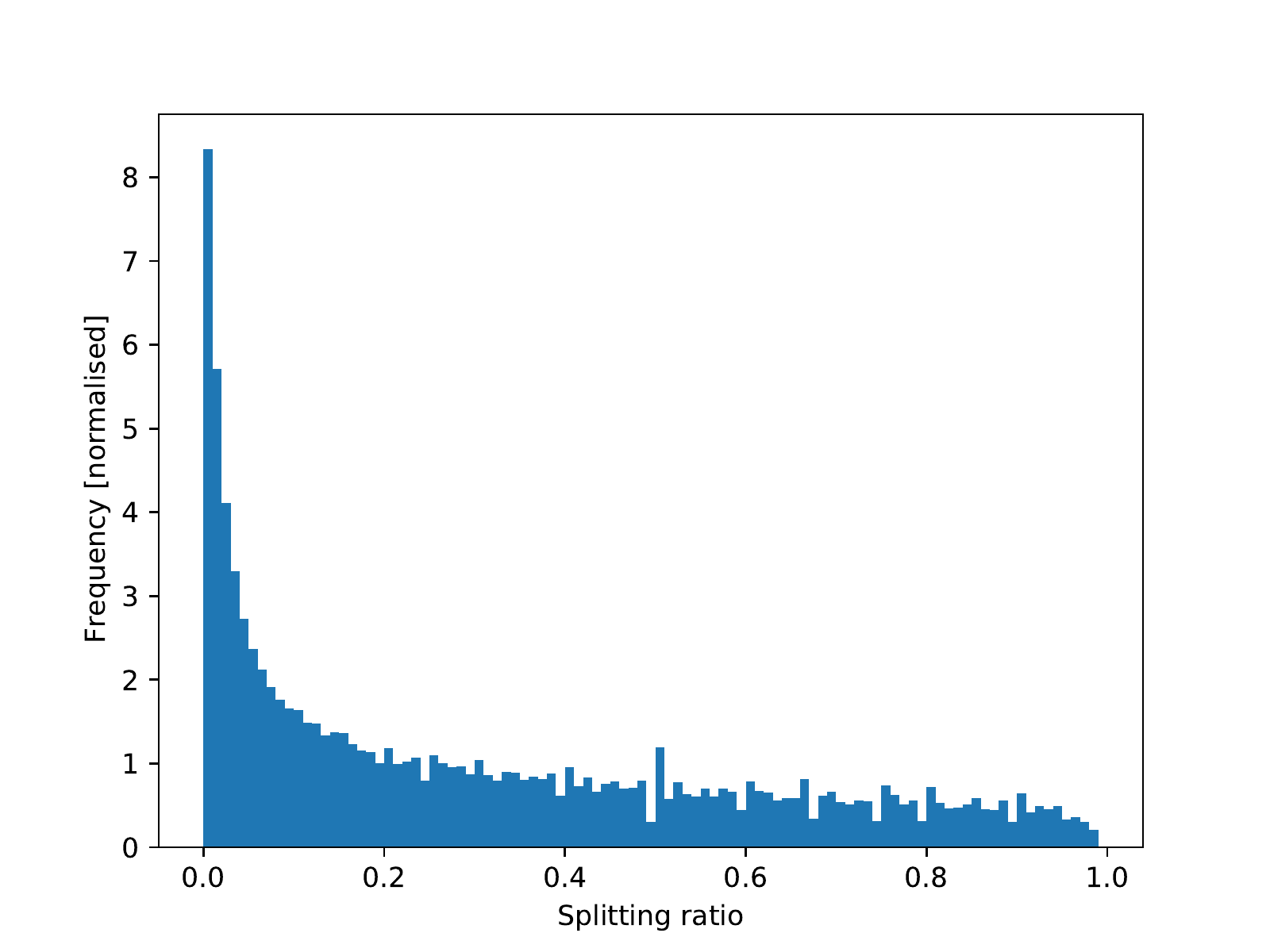}\\    
    \caption{Histograms of area ratios of fragments during the merging (top) and splitting (bottom) events for the set of forming (left) and decaying (right) active regions.}
    \label{fig:histograms_ratios}
\end{figure*}

The histogram of distances indicates that most of the merging or splitting events occur near the evolved sunspots. This is further strengthened by analysis of the area ratios of the two fragments entering the merging event or splitting into two. Histograms of these, again, separately for active regions under formation and active regions under decay, are plotted in Fig.~\ref{fig:histograms_ratios}. One can clearly see that in all cases, the area ratios are very small in general, that is, preferably, a small fragment merges with a larger one, or the large fragment splits into two, where one is much larger than the other. The latter process resembles an erosion rather than splitting.

It can also be seen that the smaller area ratios are slightly preferred in the case of the forming active regions, whereas the distribution seems wider in the case of the decaying active regions. This would indicate that during the formation, the merging of the fragments with the existing sunspots is indeed preferred. Whereas during the decay, the merging of the newly appearing fragments somewhere in the field of view is also more likely. In the decaying active regions, the number of area ratios larger than 0.5 (objects undergoing the merging or splitting events have comparable sizes) is larger by about 20\% than in the case of forming active regions.

\subsection{Correlation of quantities}
As prevously pointed out, for each investigated active region, we obtained a large set of descriptive parameters. They are best described by their evolution in time, as it is possible to see in Fig.~\ref{fig:NOAA11076_lightcurves} for an example. We consider some of these quantities important, which included the number of detected sunspots and the total object areas. We also considered the instantaneous balance between the splitting (positive) and merging (negative) events. This quantity shows whether (at the given time) merging or splitting prevails. Furthermore, we evaluated the balance between the newly appearing objects (not by splitting) and faded objects (not by merging). This quantity shows whether, at the given time, the emergence of the objects prevails (the positive value) or whether their disappearance prevails (the negative value). Finally, we recorded the total number of the detected objects at a given time. 

For every active region in the sample, it is possible to investigate the mutual relations by evaluating the Spearman's rank correlation coefficient; however, we tested to see if the use of the Pearson's correlation coefficient essentially provides the same results. We were looking to investigate whether the behaviour is somewhat common to all active regions. Thus we investigated the statistics (by histograms) of the correlation coefficients. 
\begin{figure*}
    \centering
    \makebox[0.46\textwidth]{Forming ARs} \makebox[0.46\textwidth]{Decaying ARs}\\
    \raisebox{5.5cm}{a)}\includegraphics[width=0.46\textwidth]{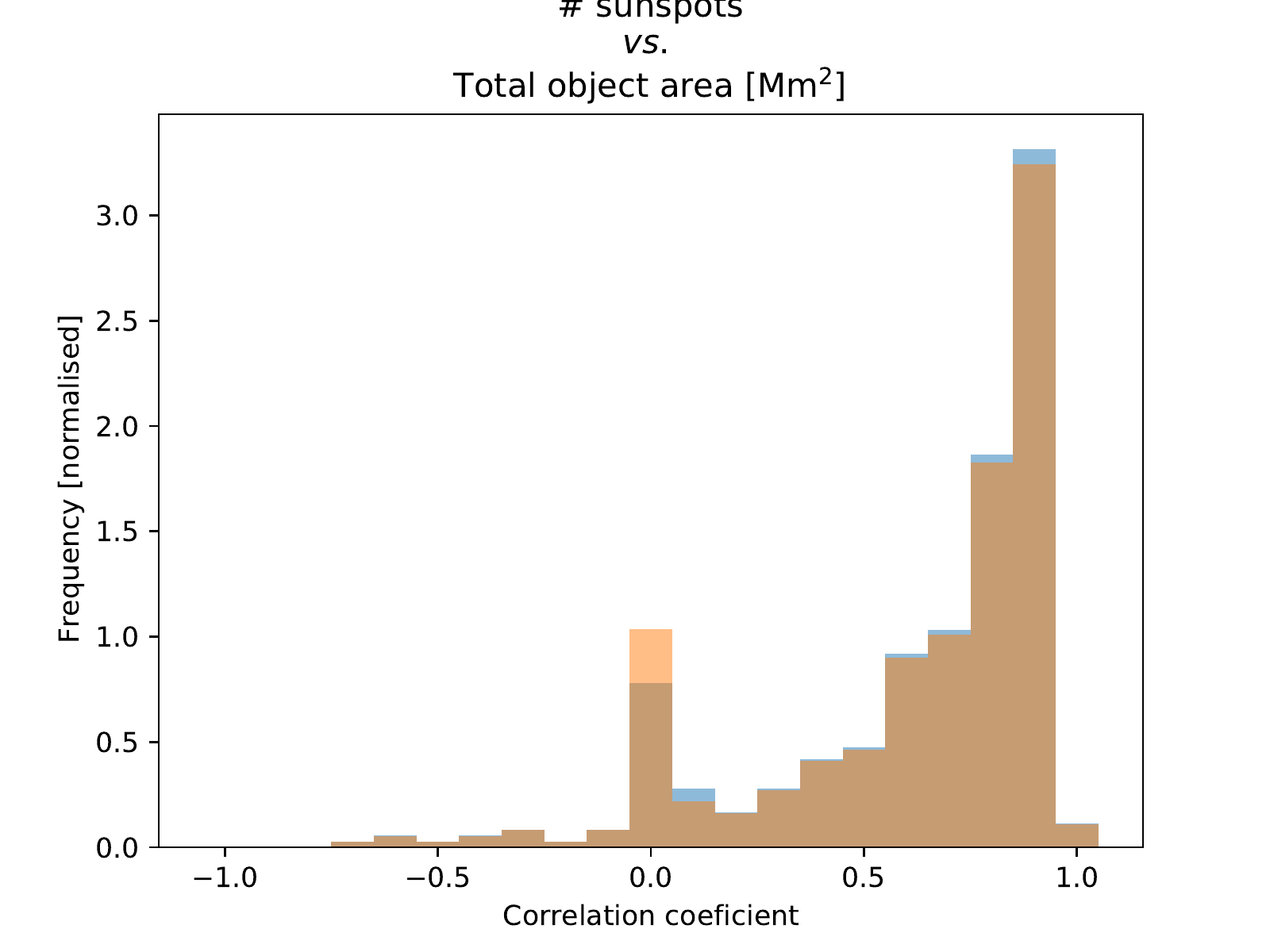}
    \raisebox{5.5cm}{b)}\includegraphics[width=0.46\textwidth]{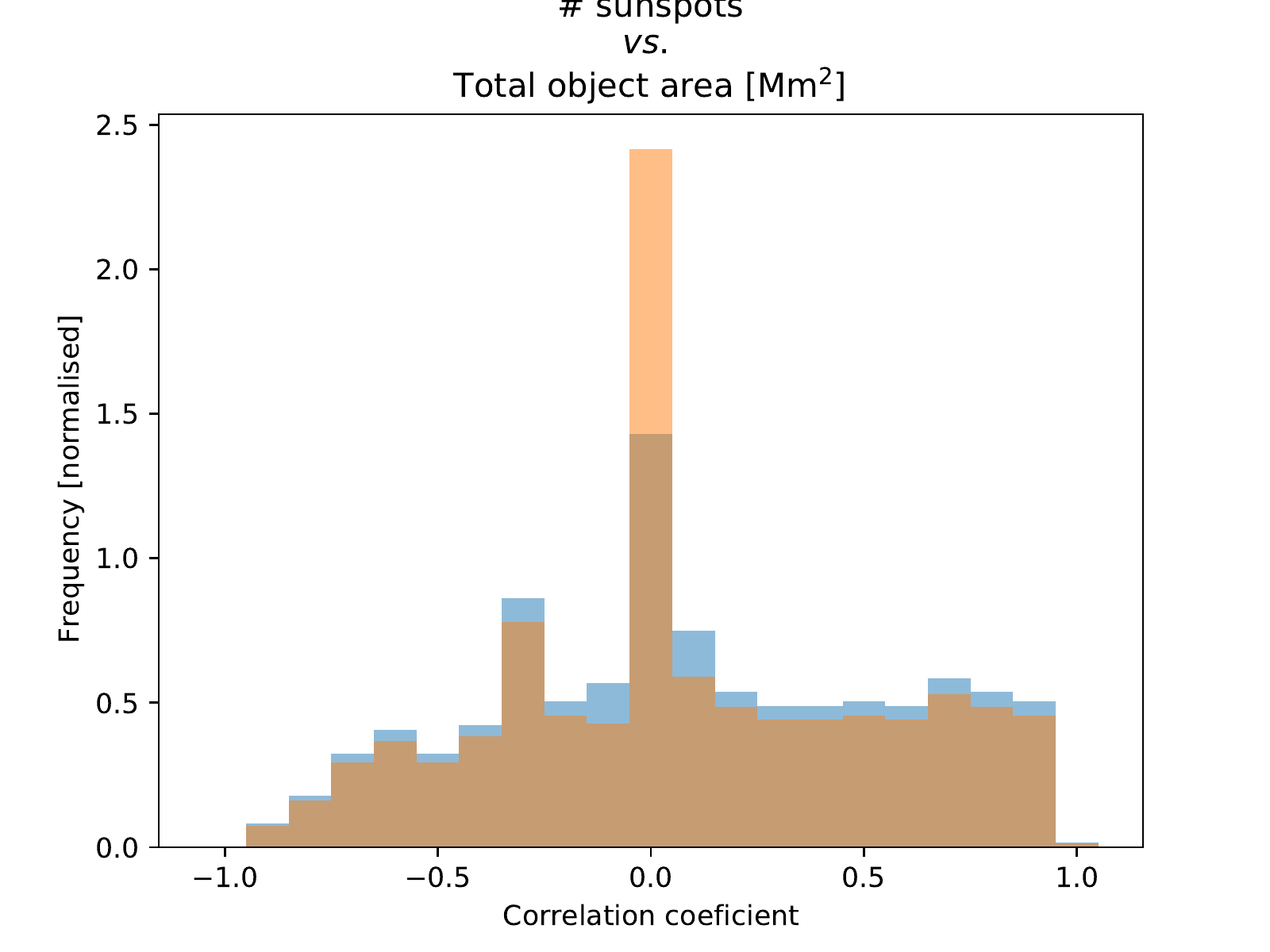}\\    
    \raisebox{5.5cm}{c)}\includegraphics[width=0.46\textwidth]{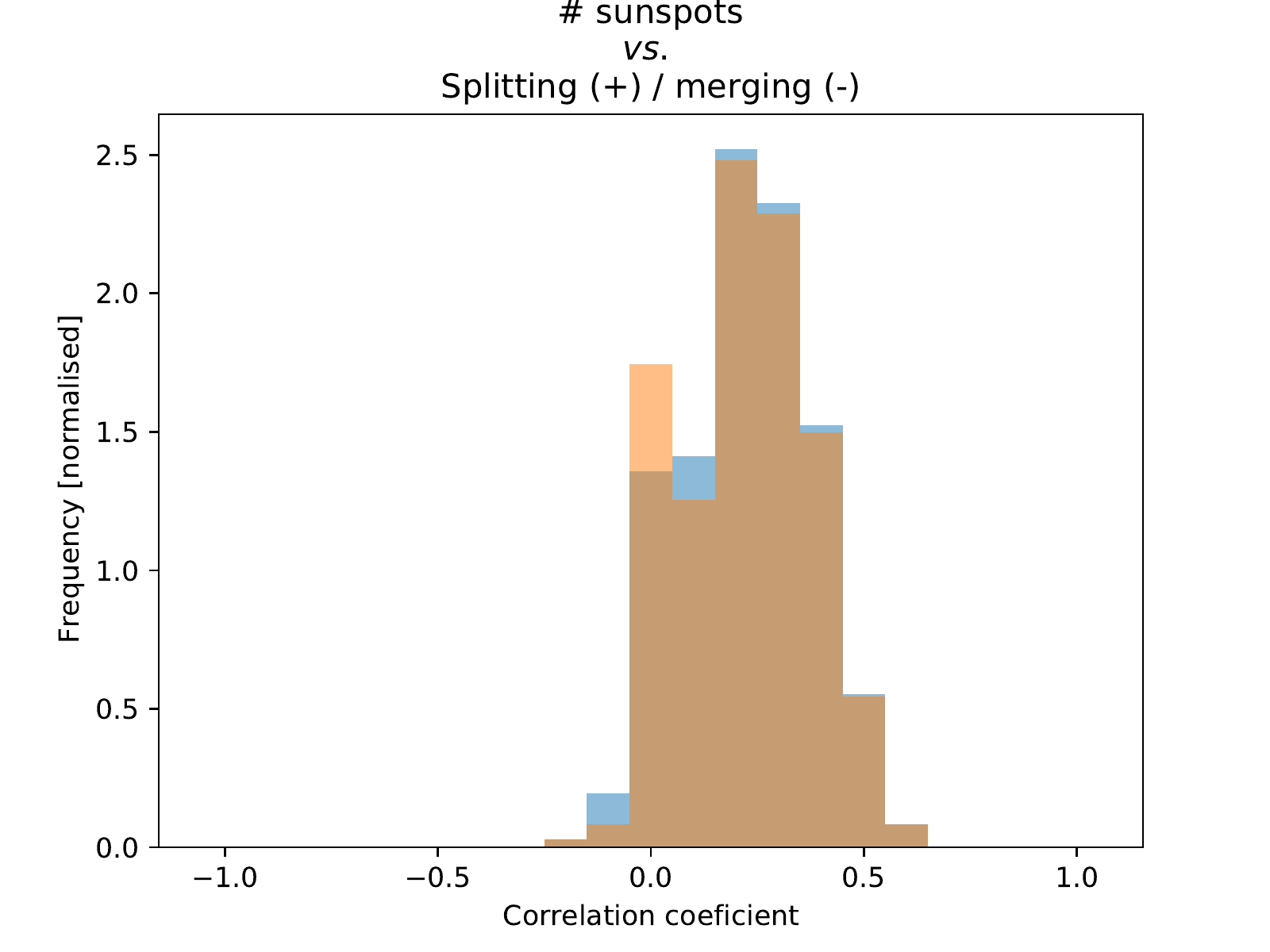}
    \raisebox{5.5cm}{d)}\includegraphics[width=0.46\textwidth]{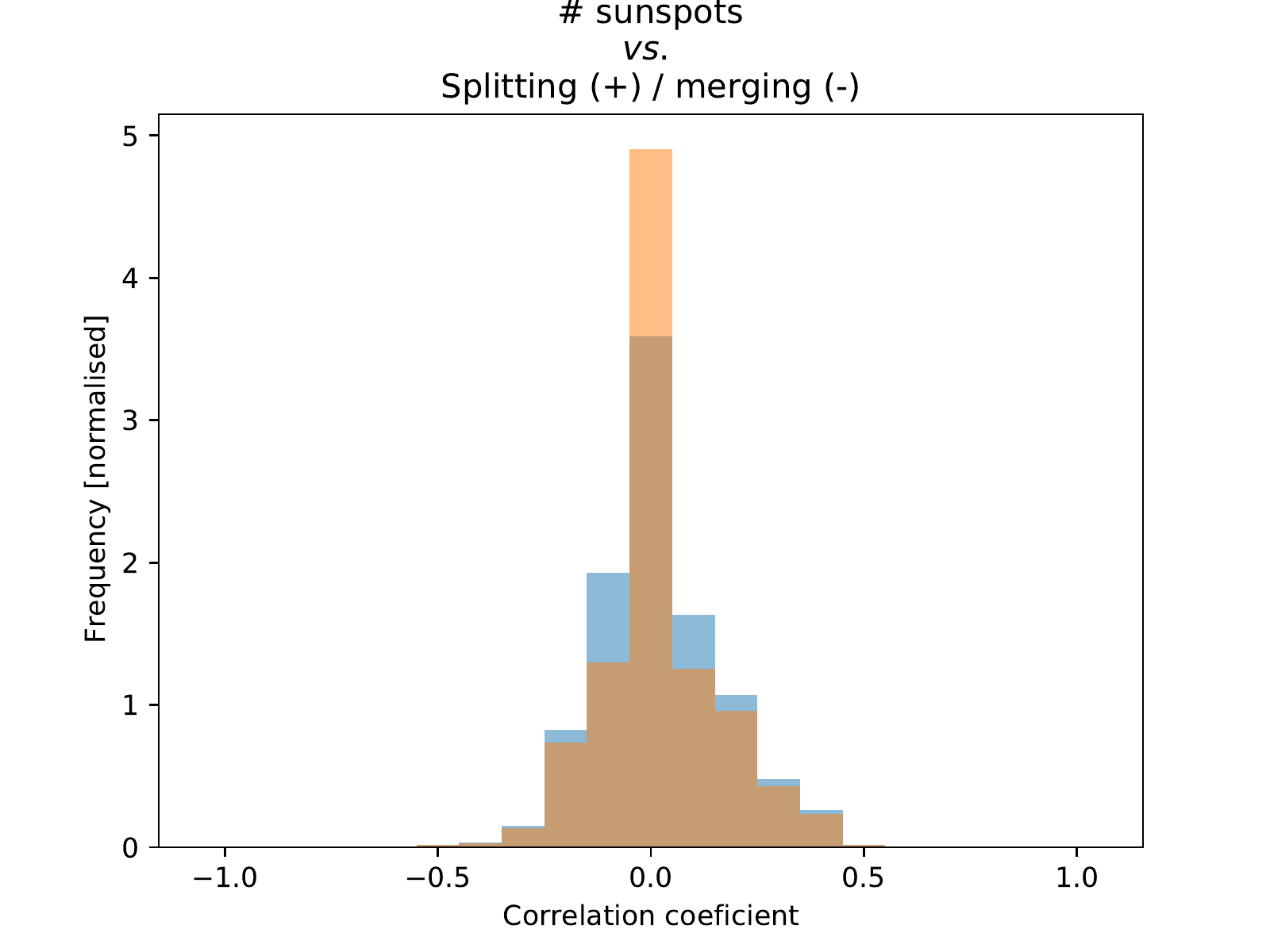}\\    
    \raisebox{5.5cm}{e)}\includegraphics[width=0.46\textwidth]{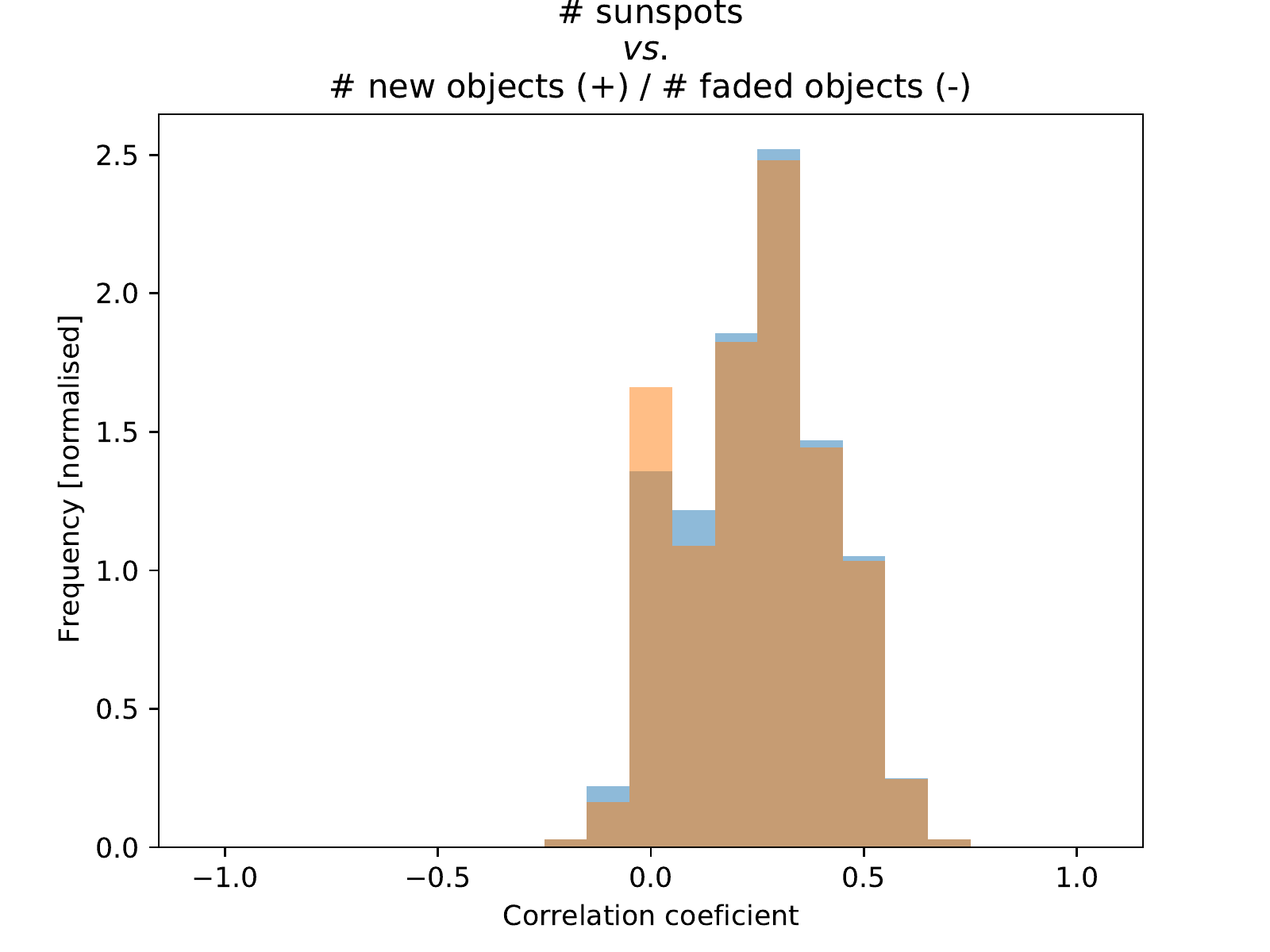}
    \raisebox{5.5cm}{f)}\includegraphics[width=0.46\textwidth]{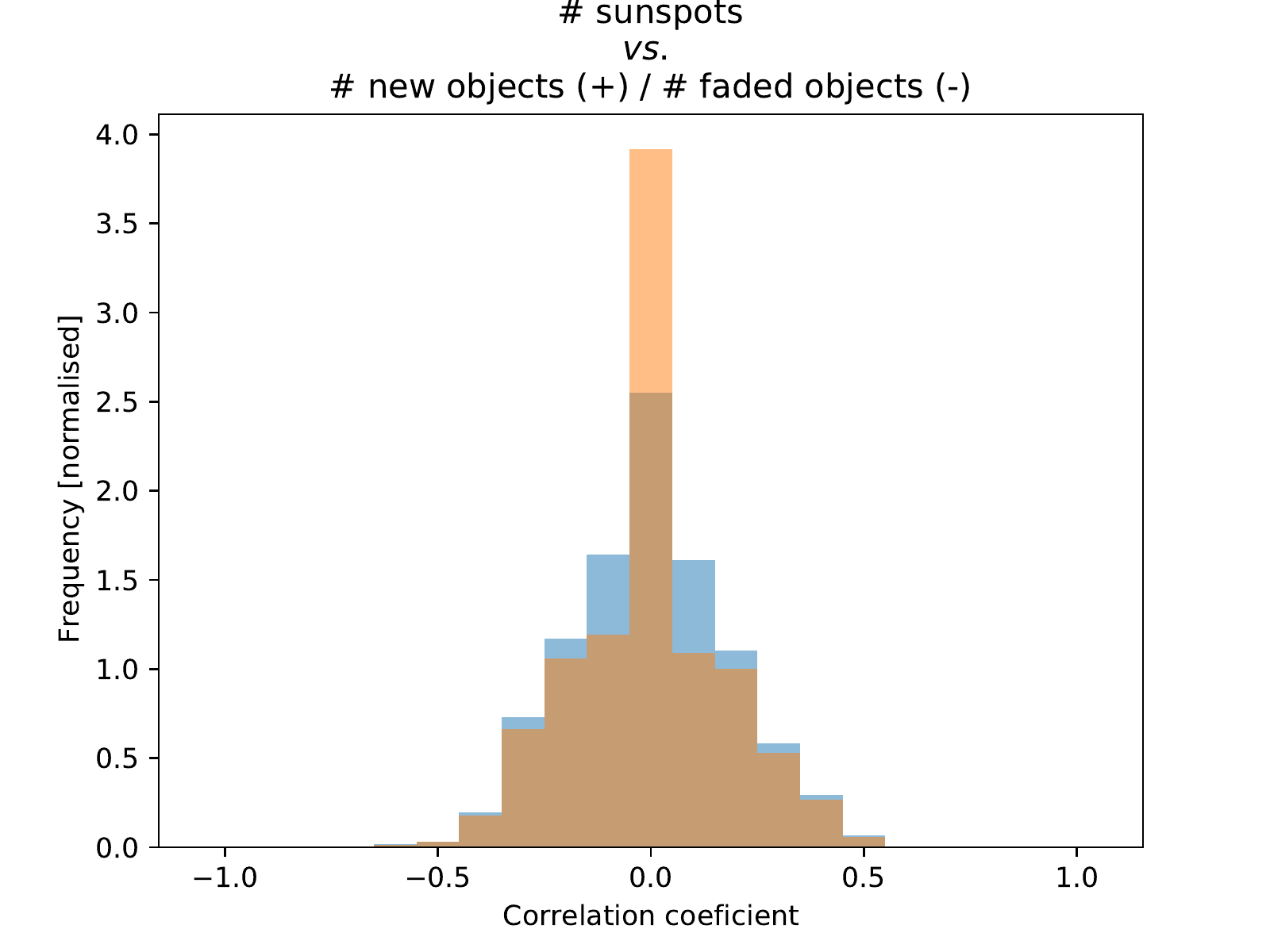}\\    
   \caption{Histograms of correlation coefficients between various quantities describing the evolution of the fragments in the forming (left) and decaying (right) active regions. The blue bars represent the the histograms of all correlation coefficients, whereas the orange-bar histograms were derived using only the statistically significant correlations coefficients. The brown colour indicates the overlap of both types of histograms.}
    \label{fig:correlations}
\end{figure*}

\begin{figure*}
    \centering
    \makebox[0.46\textwidth]{Forming ARs} \makebox[0.46\textwidth]{Decaying ARs}\\
     \raisebox{5.5cm}{g)}\includegraphics[width=0.46\textwidth]{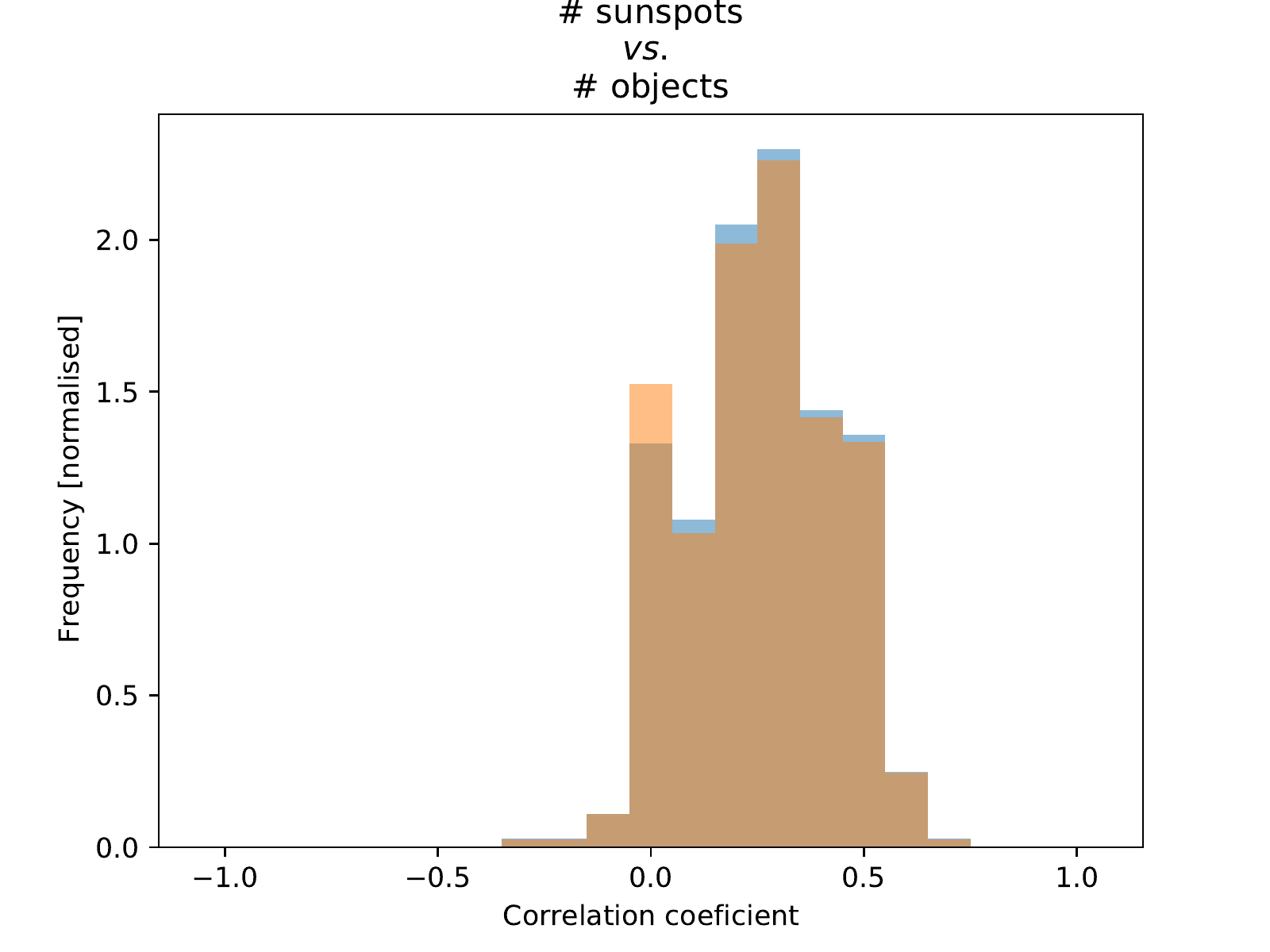}
    \raisebox{5.5cm}{h)}\includegraphics[width=0.46\textwidth]{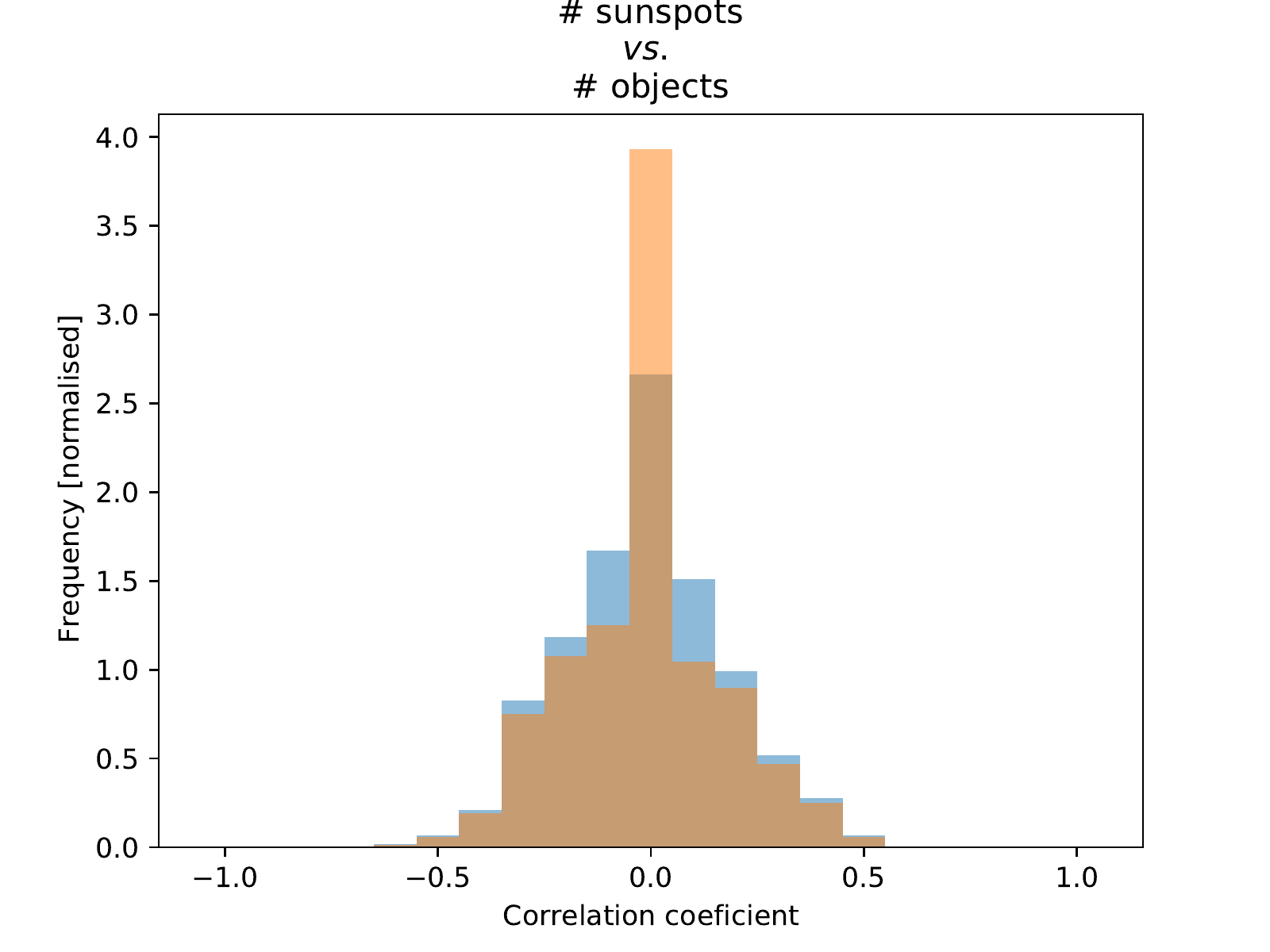}\\    
    \raisebox{5.5cm}{i)}\includegraphics[width=0.46\textwidth]{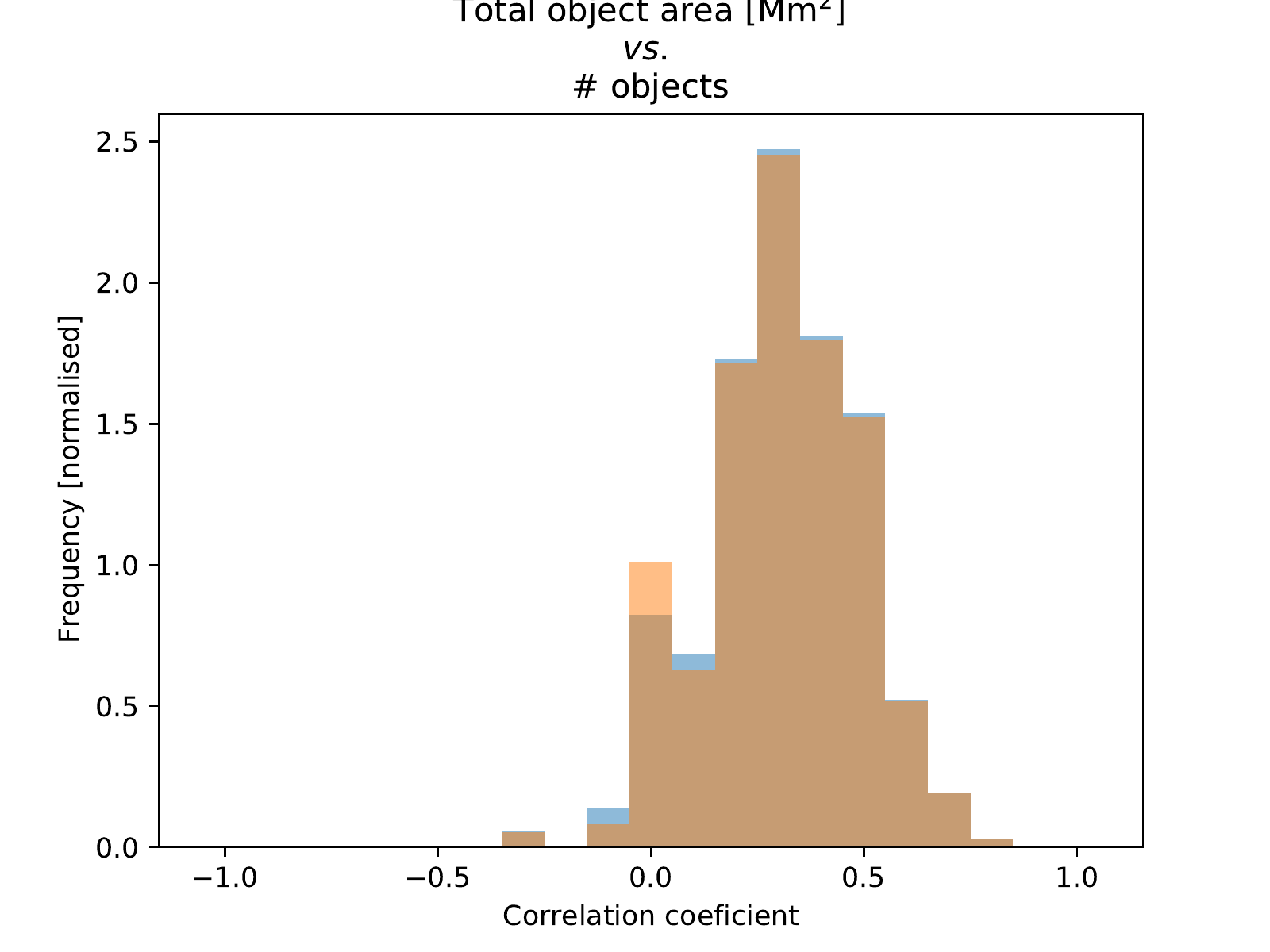}
    \raisebox{5.5cm}{j)}\includegraphics[width=0.46\textwidth]{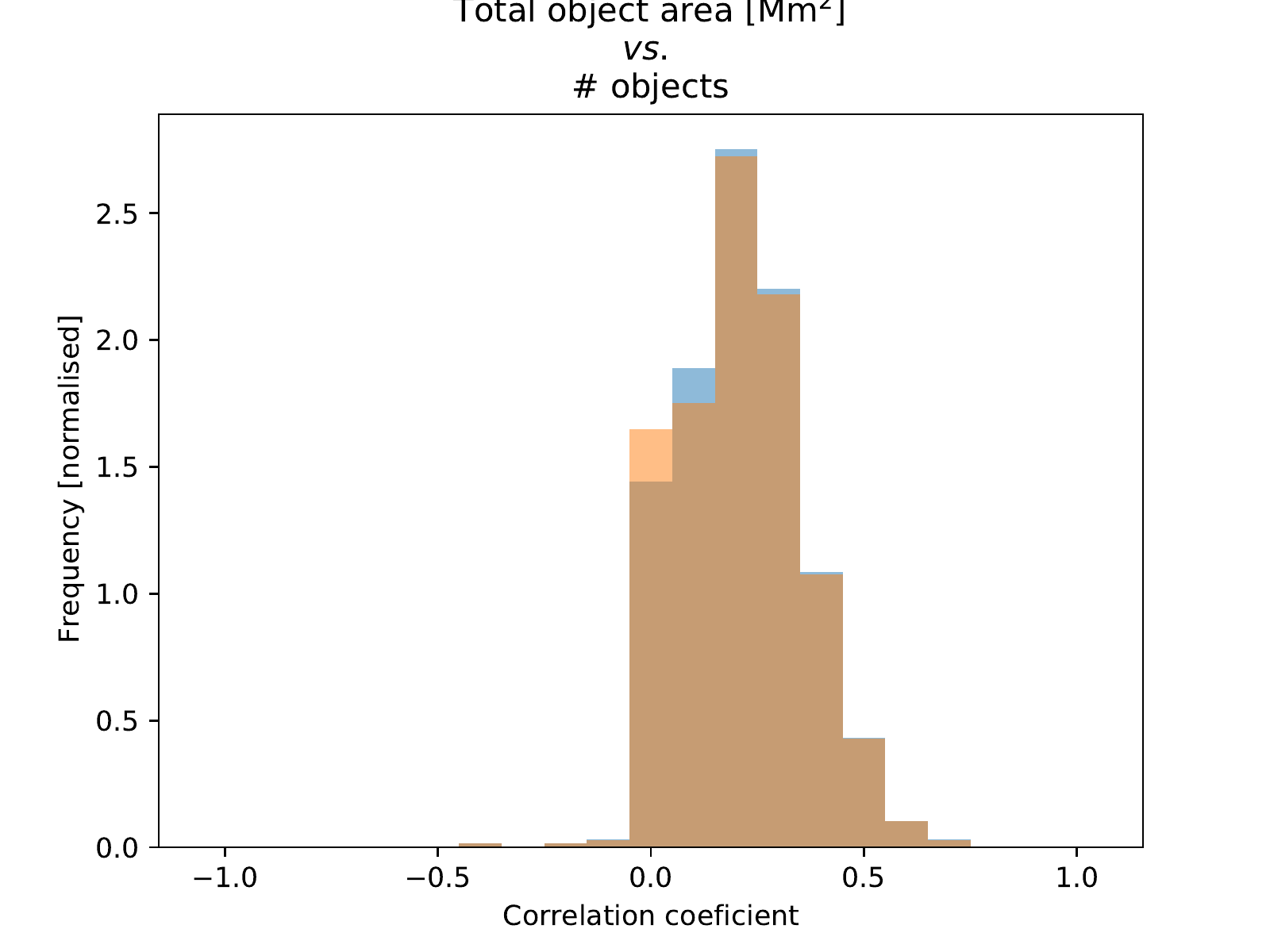}\\   
    \raisebox{5.5cm}{k)}\includegraphics[width=0.46\textwidth]{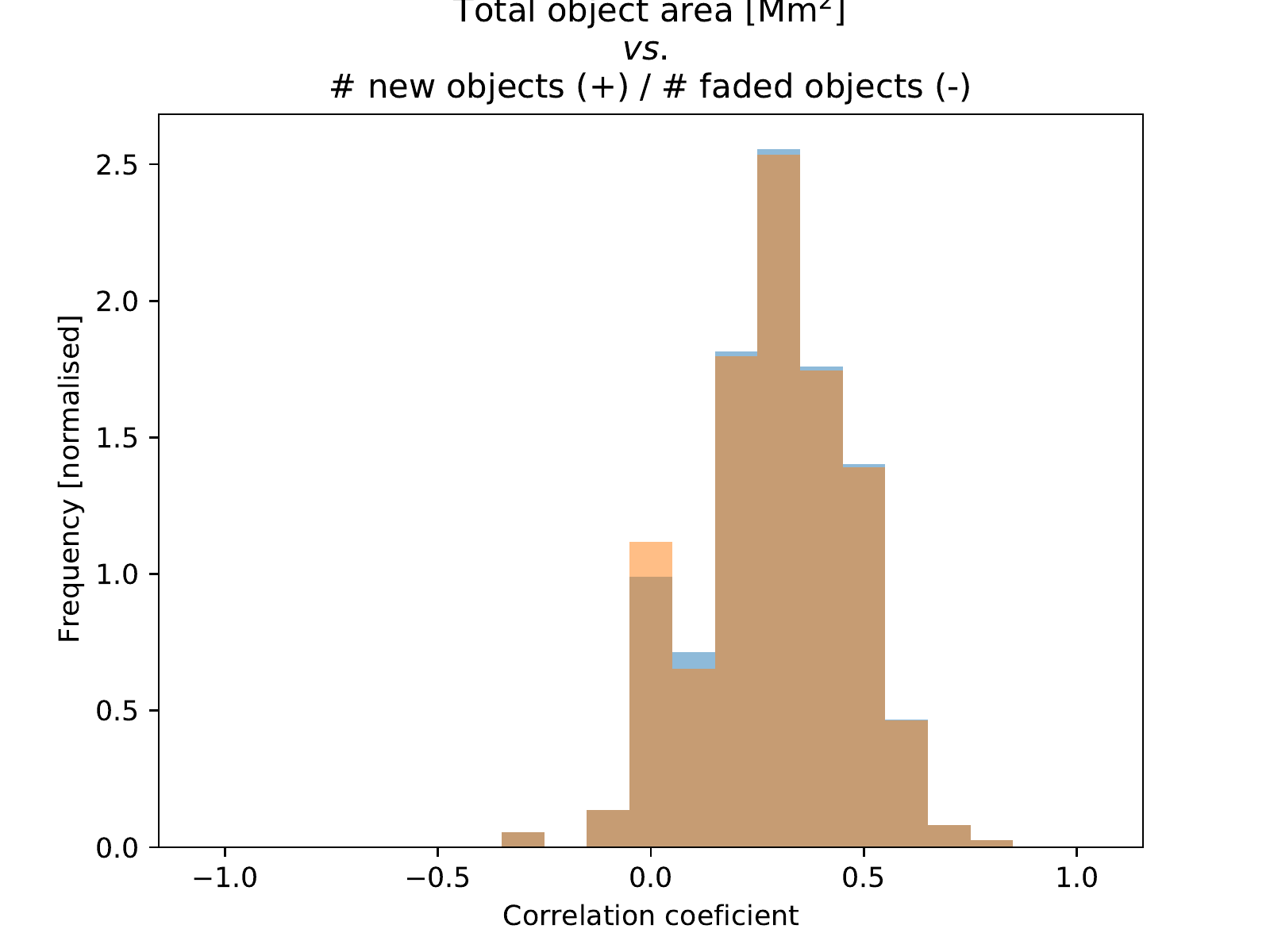}
    \raisebox{5.5cm}{l)}\includegraphics[width=0.46\textwidth]{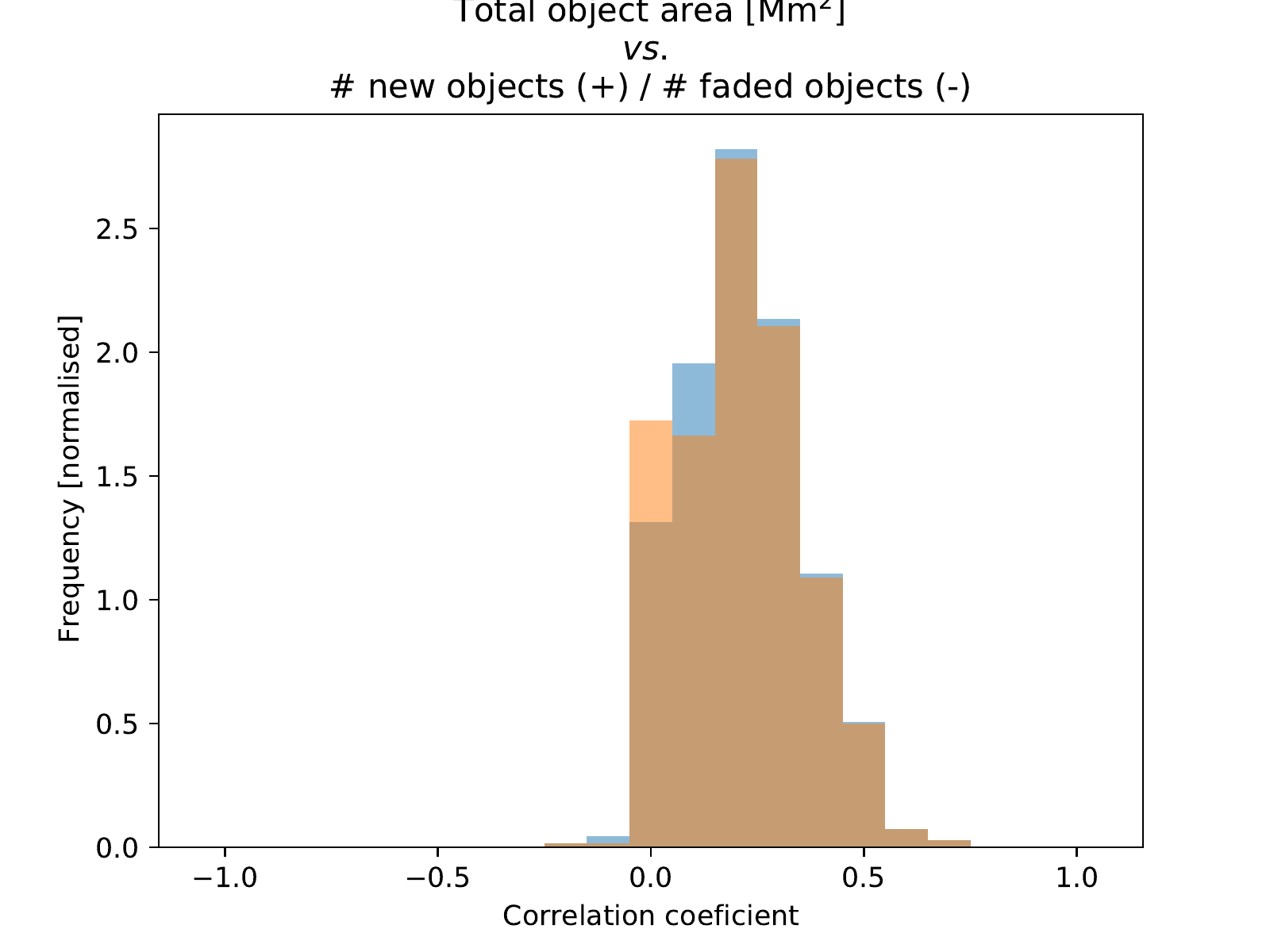}\\    
    \addtocounter{figure}{-1}
    \caption{Continued}
\end{figure*}

\begin{figure*}
    \centering
    \makebox[0.46\textwidth]{Forming ARs} \makebox[0.46\textwidth]{Decaying ARs}\\
    \raisebox{5.5cm}{m)}\includegraphics[width=0.46\textwidth]{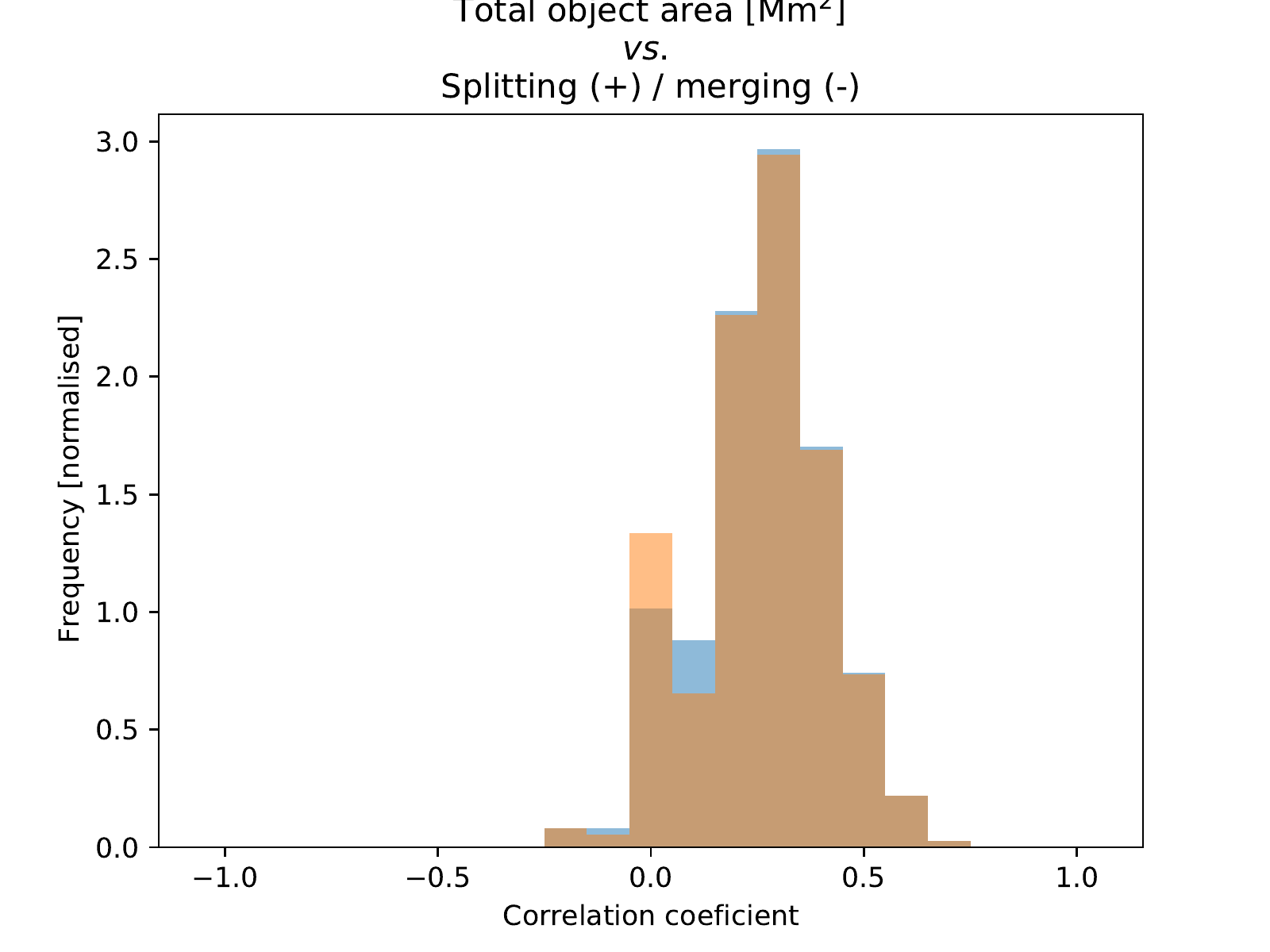}
    \raisebox{5.5cm}{n)}\includegraphics[width=0.46\textwidth]{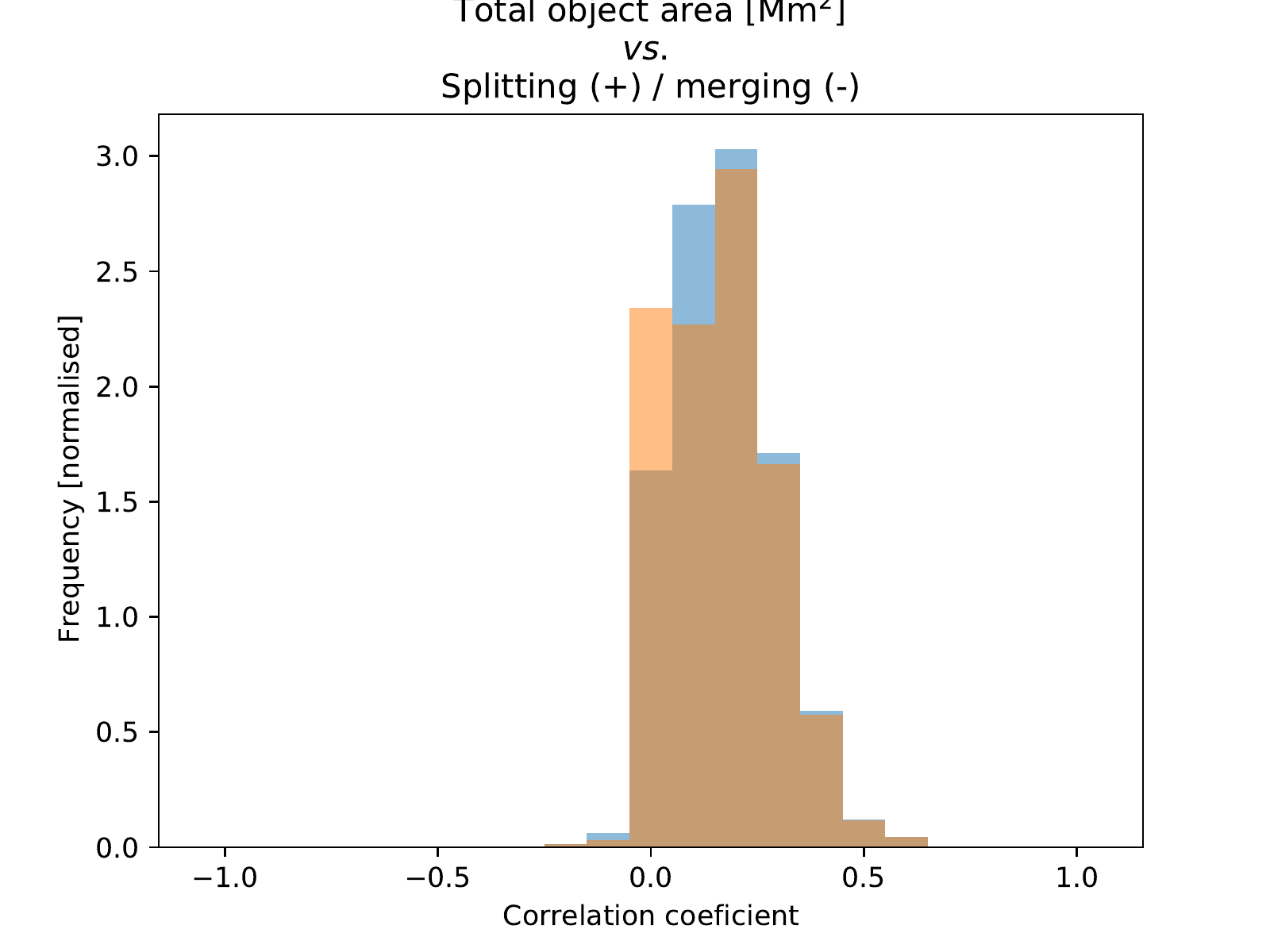}\\    
    \raisebox{5.5cm}{o)}\includegraphics[width=0.46\textwidth]{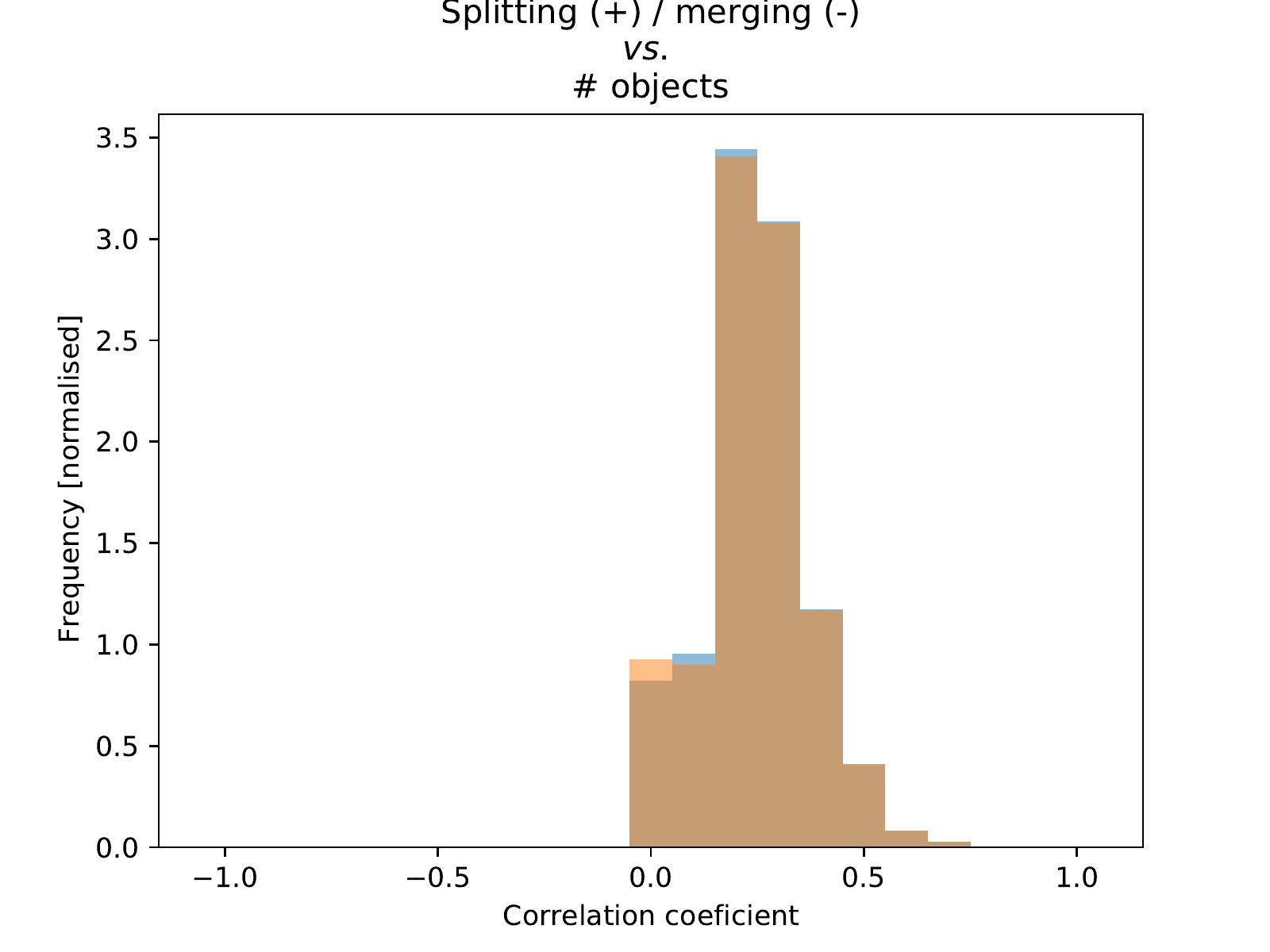}
    \raisebox{5.5cm}{p)}\includegraphics[width=0.46\textwidth]{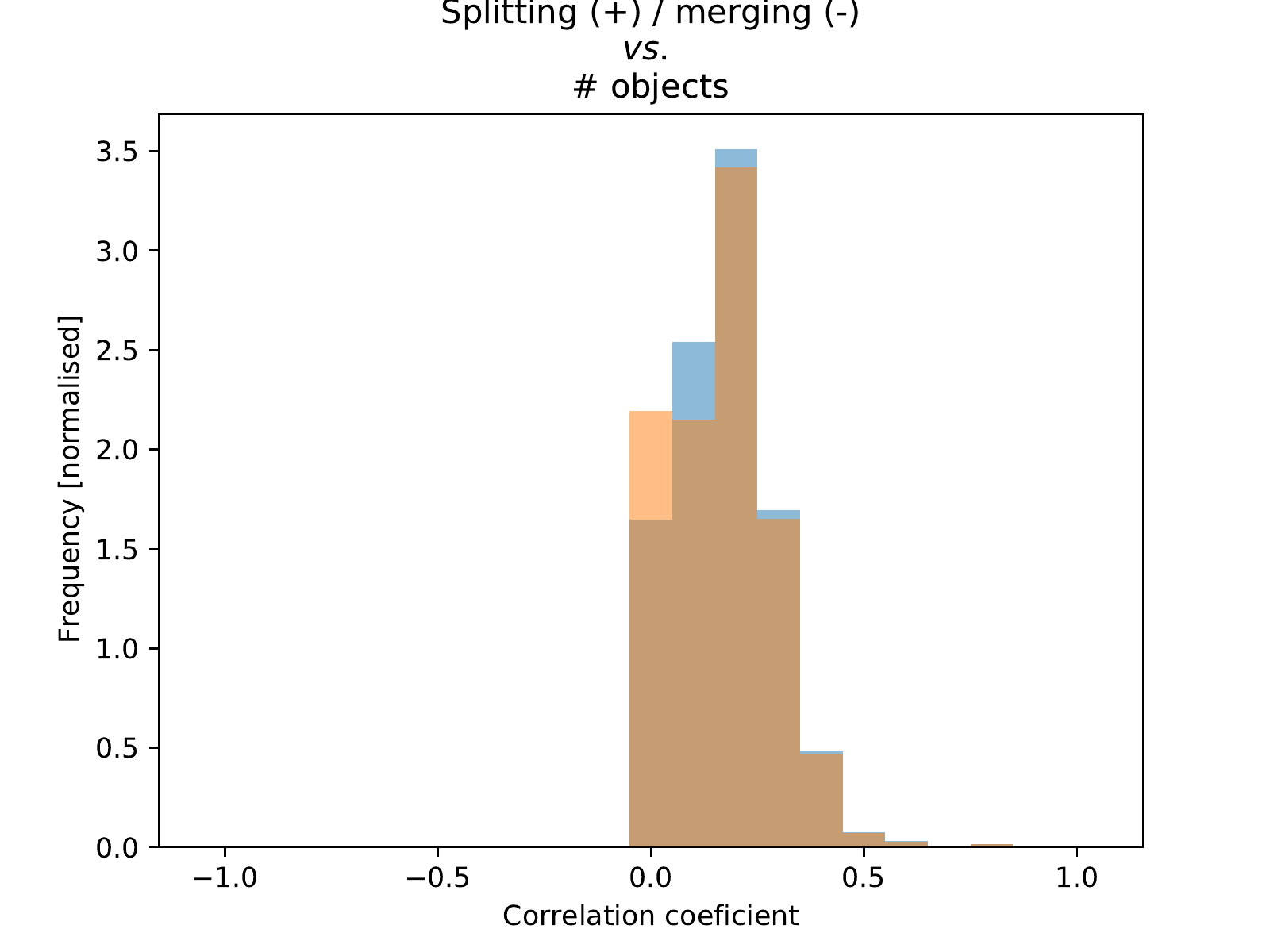}\\    
    \raisebox{5.5cm}{q)}\includegraphics[width=0.46\textwidth]{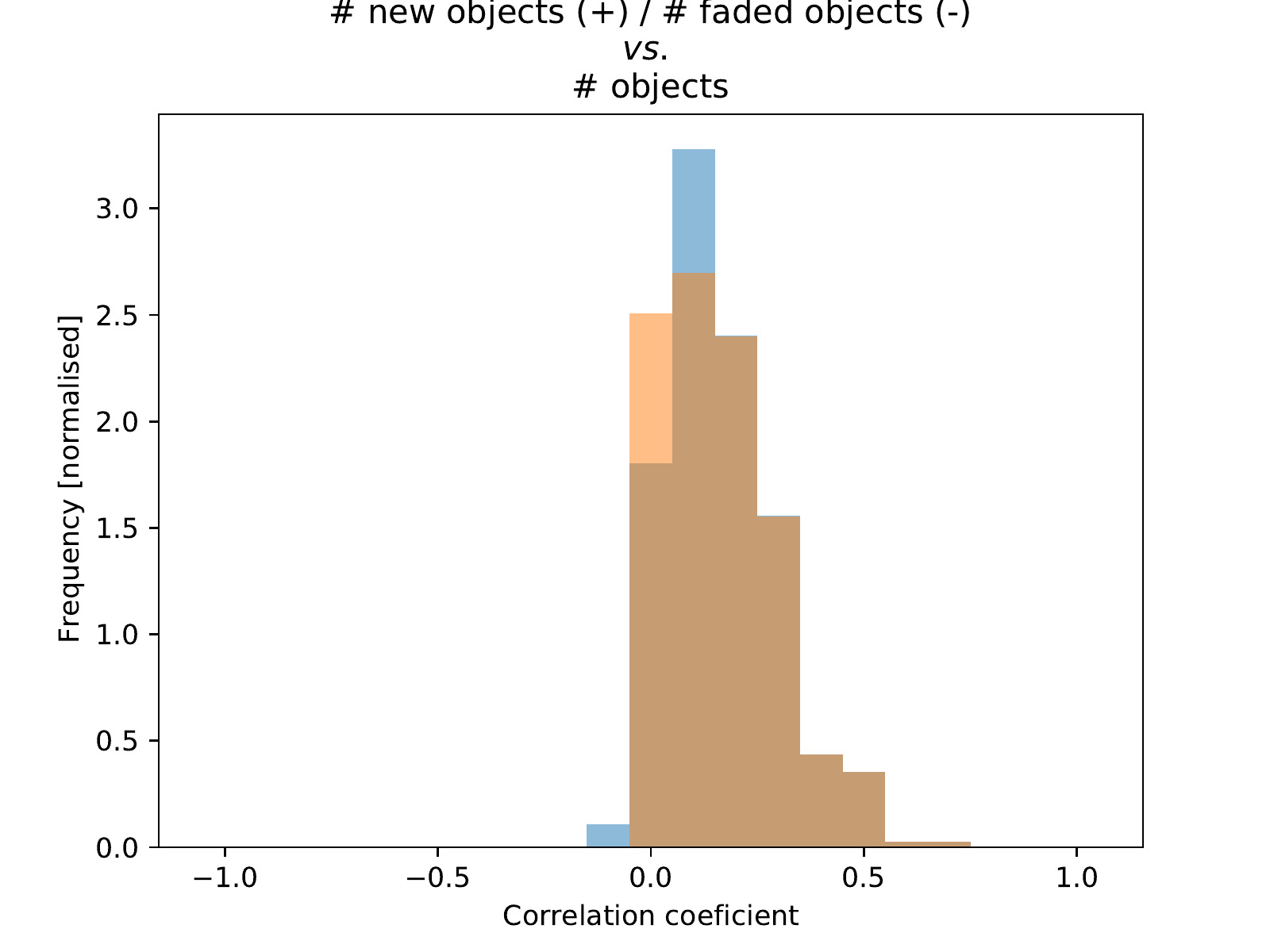}
    \raisebox{5.5cm}{r)}\includegraphics[width=0.46\textwidth]{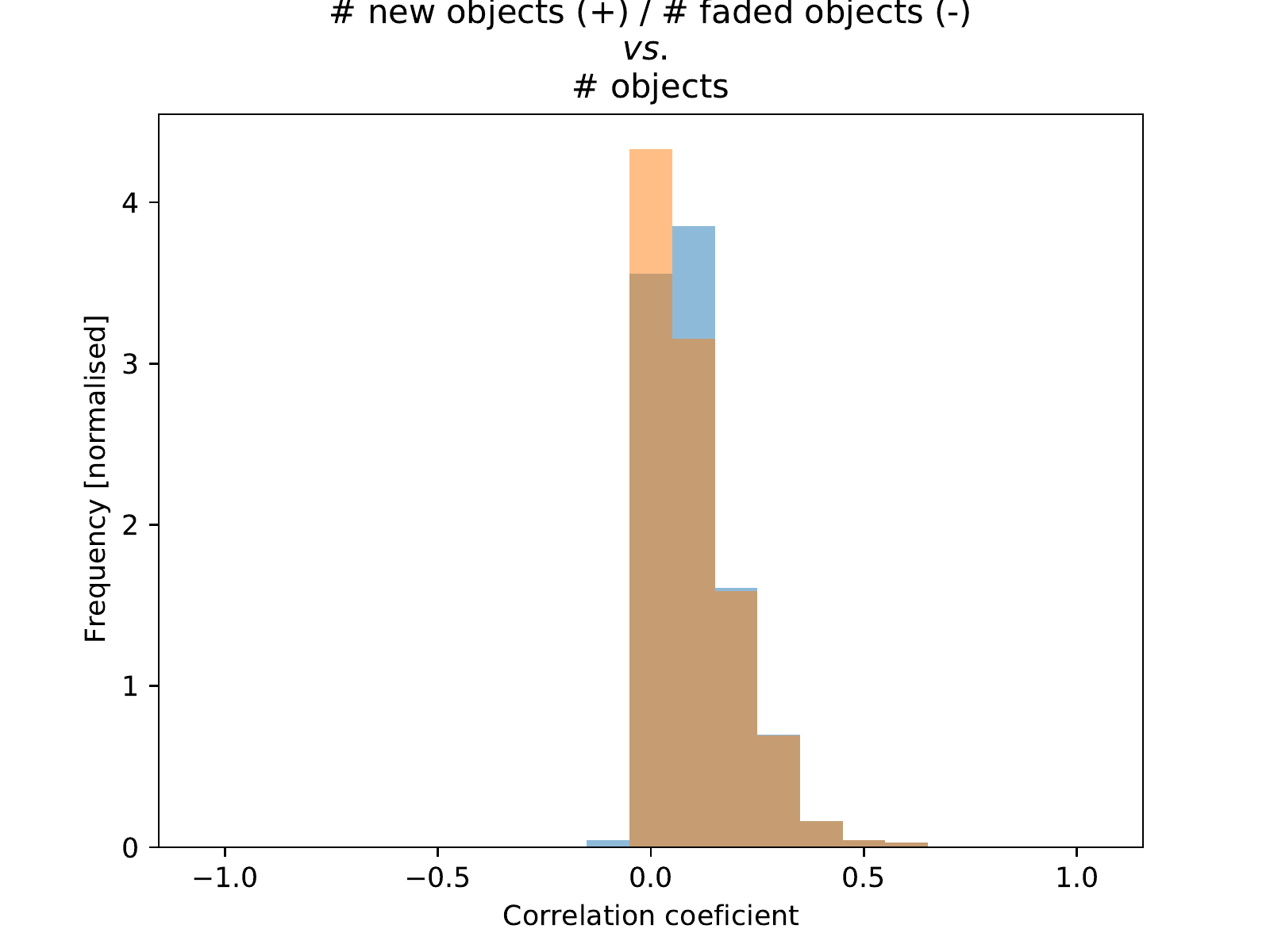}\\   
    \addtocounter{figure}{-1}
    \caption{Continued}
\end{figure*}

The histograms of the correlations of relevant quantities are given in Fig.~\ref{fig:correlations}. They are plotted in two versions: The blueish bars aggregate all the correlation coefficients that were computed separately for each active region in the sample and the orangish bars only aggregate those correlation coefficients, whose $p$-values were smaller than the significance threshold (0.05). The $p$-value gives the probability that the correlation indicated by the correlation coefficient is only apparent and the two correlated series are in fact independent. Hence, in the case when the $p$-value was larger than the threshold, the null testing hypothesis of non-correlation of the series was taken into account and the value of the correlation coefficient was thus set to 0.0 for the given pair. As one can see, the differences between the two approaches are not significant. The histograms were normalised to the area so that the area integral (or sum strictly speaking) of the histograms over all bins yields unity. 

The displayed correlations seem to behave similarly for both sets of forming and decaying active regions with the exception of the correlation of the quantities with the number of sunspots. In the case of the forming active regions, positive or slightly positive correlations prevail; while in the case of the decaying active regions, the correlations group around zero. 

For instance, the relation between the number of sunspots and the total object area is clearly shifted towards large positive correlations (Fig.~\ref{fig:correlations} a). During the emergence, new sunspots appear and grow in size. The growing objects are classified as sunspots eventually. On the other hand, for the decaying active regions, the correlation takes all values with a strong peak around zero (Fig.~\ref{fig:correlations} b). This could indicate that during the decay, sunspots likely fragment to smaller pieces. Their total area decreases as well but on a longer timescale than the typical splitting timescale. Hence the correlation between the sunspot number changes, which indicates the evolutionary state of the active regions, and the change of the total area shows a large peak around zero and a broad distribution in the values.  

The cumulative curve of the splitting and merging events in the case of the forming active regions is dominated by merging events (see Fig.~\ref{fig:NOAA11076_lightcurves} upper right panel). The prevailing rather positive correlation between the instantaneous splitting and merging curve and the number of sunspots (Fig.~\ref{fig:correlations} c) is dominated by sunspot merging in the forming phase of the active region. Many fragments in this phase have already been classified as sunspots (see a further discussion of Fig.~\ref{fig:correlations} g in the next paragraph) and thus the number of merging events correlates with a decrease in the number of sunspots. This is consistent with the work by \cite{2019SoPh..294..102K}, where they point out that during the emergence, the coalescence is an important process; however, the emergence of large structures brings more energy to the forming active region. In the case of the decaying active regions, the correlation between the same quantities is zero (Fig.~\ref{fig:correlations} d). The splitting and merging curve is dominated by splitting events in this case. However, the number of sunspots does not increase accordingly. This may be understood in a way that the splitting events result in small objects that are not classified as proper sunspots. This again resembles an erosion process rather than splitting into two comparable fragments. 

The number of sunspots and the curve of newly appearing (not by
splitting) and fading (not by merging) objects (Fig.~\ref{fig:correlations} e,f) are positively correlated in the case of forming
active regions.
This is consistent with the emergence of new objects dominating their spontaneous disappearance during the emerging phase of the active region when the number of identified sunspots grows in general. In the case of decaying active regions, the imbalance between the newly appearing and fading objects has nothing to do with the actual number of remaining sunspots in the area. This is consistent with a random process, which is further supported by the distribution of distances between the merging and splitting events and the sunspots. This distribution of distances is wider in the case of decaying active regions (Section~\ref{sect:locations}). If the active region in the decaying phase is already disconnected from its magnetic roots \citep[suggested e.g. by][]{1994ApJ...436..907F,2005AA...441..337S}, the emergence of the new dark objects must be due to the dynamo process operating in the near-surface layers.  
A similar observation is drawn about the correlation between the number of the detected sunspots and the number of objects (Fig.~\ref{fig:correlations} g,h). The weak positive correlation for the forming active regions has two interpretations. Either a considerable fraction of objects already emerge as objects that are large enough to be classified as sunspots, or these objects emerge with the properties below the sunspot classification but coalesce quickly with the other objects and become sunspots. We remind our readers that the cadence of our dataset is 12 minutes.  
Both interpretations are consistent with the observation that often, when an object is registered for the first time, it already has the properties of sunspots. To shed some light on the discrimination between these two hypotheses, one would need to study the datasets with a higher temporal resolution, for instance with the origin 45-s sampling of the HMI. Unfortunately, such a test is currently beyond our computer capabilities. In the case of decaying active regions, the number of objects has nothing to do with the number of detected sunspots. 

The total area covered by the objects weakly correlates with the total number of identified objects (Fig.~\ref{fig:correlations} i,j). The distribution of the correlation coefficients is similar for both the forming and decaying active regions; however, it is slightly shifted towards larger values in the case of the forming active regions. This weak correlation indicates that the sizes of the objects differ and no typical size exists. If a typical size existed, we would expect a clear (even linear) correlation between these two values. The same conclusion may be drawn about the correlation between the total area and the imbalance between the newly appearing and spontaneously disappearing objects (Fig.~\ref{fig:correlations} k,l).

Something that is somewhat surprising is a weak positive correlation between the total area covered by the objects and the imbalance between the number of splitting and merging events (Fig.~\ref{fig:correlations} m,n). One would naively expect that during the splitting and merging, the area is conserved. Thus when the splitting occurs, the splitting and merging curve increases, but the total area should stay the same size. When the merging occurs, the splitting and merging curve decreases, but the total area should remain constant again. Therefore the expected correlation coefficient value is zero. The positive values of correlation indicate that the summary area of the resulting fragments is larger than the area of the mother object and, similarly, that for the merging event, the merger has a smaller area than the sum of the areas of the predecessor objects. If we assume that the total magnetic flux is conserved during the merging and splitting events, we may say that the average magnetic induction of the objects decreases after splitting and increases after merging.  

Unfortunately, a finite spatial resolution may play a significant role here. The HMI resolution will likely lead to an overestimation of the areas of smaller fragments, against the area of bigger fragments. The bias will be relatively larger for very small fragments, as the accuracy of measuring the area is higher when the area is significantly larger than the pixel size. A coarse grid of pixels may lead to a relatively lower intensity of small structures, as a result of smaller filling factor, but to an overall larger area, unless the contrast is low enough to inhibit its detection. Therefore the positive correlations seen in Fig.~\ref{fig:correlations} m,n may be a consequence of the sampling. 

The splitting and merging and newly appearing and spontaneously disappearing curves weakly correlate with the total number of objects (Fig.~\ref{fig:correlations} o,p and q,r). This indicates that neither of the processes is dominant in any phase of the active region lifetime. The correlation seems a bit larger for the splitting and merging curve, showing slightly greater importance of this process on the total number of detected objects.

\subsection{Speeds of the merging fragments}
We now focus on merging events, namely on the evolution of the fragment between its appearance and its merging with another fragment. We have shown that in most cases, it is the smaller fragments that merge with the larger ones. Their lifetime is rather small, a few hours at most, but most often it is less than one hour (i.e. five frames taking the cadence of our data into account). 

\begin{figure}
    \centering
    \includegraphics[width=0.46\textwidth]{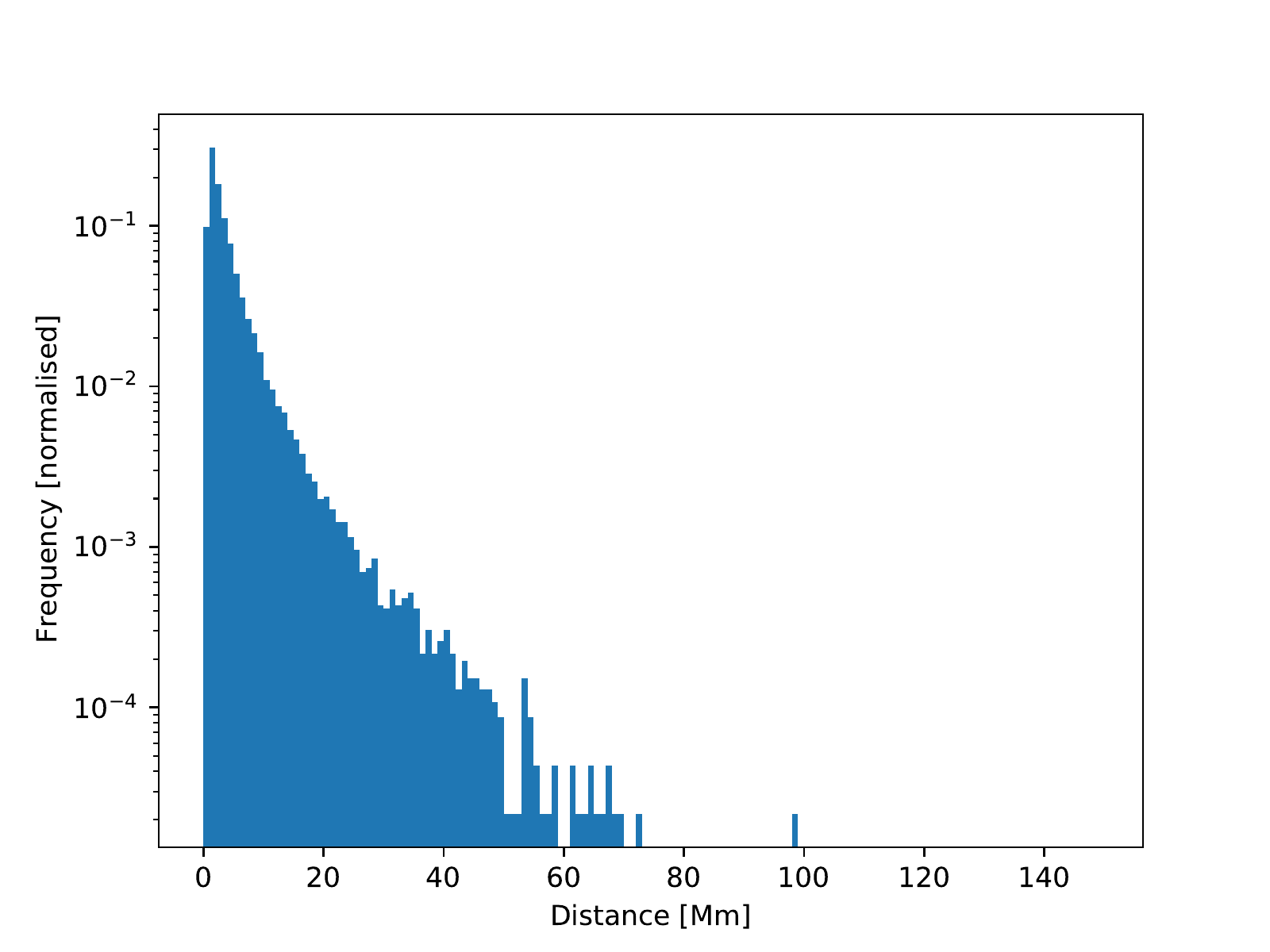}
    \caption{Histogram of maximal distances of fragments during its lifetime from the position where it merged with another one. It is important to note the logarithmic scaling on the vertical axis.}
    \label{fig:hist_maxistances}
\end{figure}

We studied the statistics of the maximum distance during the object lifetime and the position of its death by merging. Fig.~\ref{fig:hist_maxistances} shows that in most cases, this distance is rather small with a maximum at about 2~Mm.

\begin{figure}
    \centering
    \includegraphics[width=0.46\textwidth]{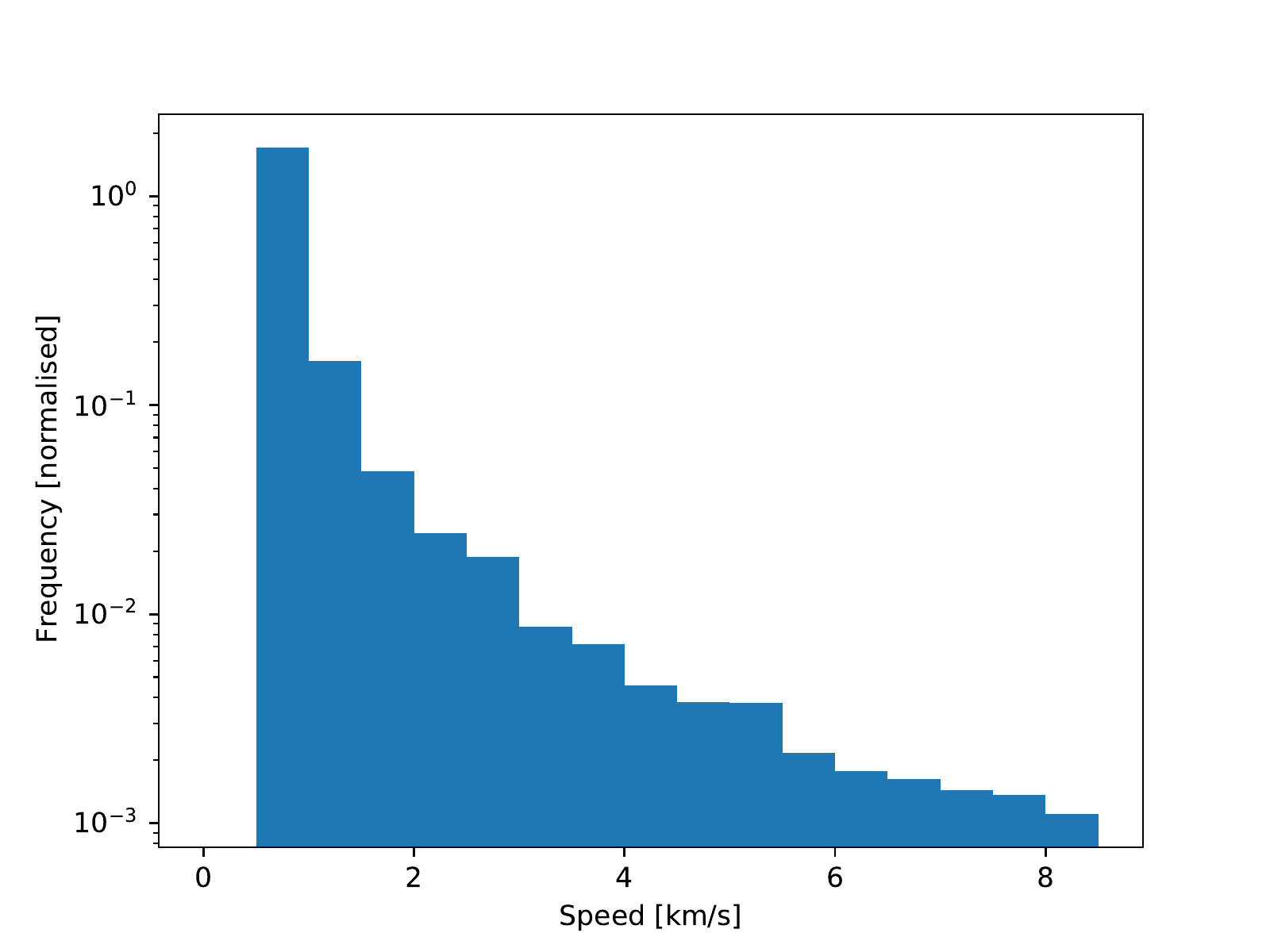}
    \caption{Histogram of speeds of the fragments approaching the merging event. We note the logarithmic scaling on the vertical axis.}
    \label{fig:hist_speeds}
\end{figure}

We also investigated, how fast do the fragments approach each other. Fig.~\ref{fig:hist_speeds} indicates that these speeds are rather slow, about 0.5~km/s, which also corresponds to the minimum velocity detectable given the cadence and sampling of the data. We have to add that null velocities were removed from this plot. The speed of 0.5~km/s corresponds to the typical speed of the supergranular diffusion. In studies performed using high-resolution observations, such as \cite{2010ApJ...724.1083G} for example, higher speeds were observed for small magnetic elements, where the recorded motions were dominated by granular flows, which may be an order of magnitude larger. Such small scales are out of reach when using HMI/SDO measurements. 

\begin{figure}
    \centering
    \includegraphics[width=0.46\textwidth]{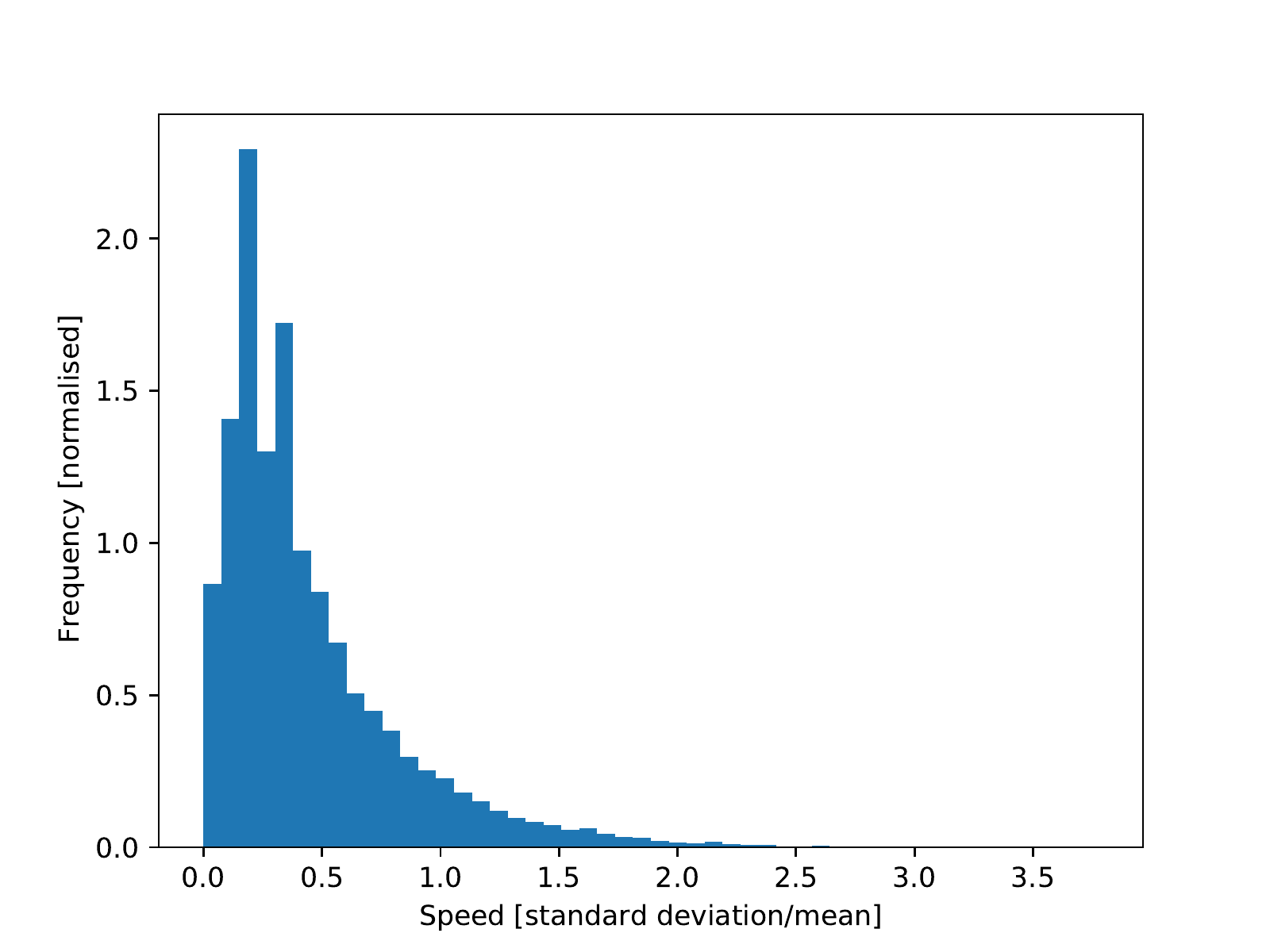}
    \caption{Histogram of speed standard deviations expressed as the fraction of the mean speed of the fragments approaching the merging event.}
    \label{fig:hist_norm_speeds}
\end{figure}

We studied, how does this velocity vary during the motion. Fig.~\ref{fig:hist_norm_speeds} indicates that the speed does not vary much as the distribution of the speed standard deviations that are normalised by the mean speeds is a heavy-head distribution. This indicates that the approaching speed towards the merger of the fragments is more or less constant during the whole process. 

We were interested to see, whether the fragments go straight towards the merging centre or whether this process is rather similar to the random-walk process. The instant directions towards the merger positions are defined by the angle $\varphi$ as
\begin{equation}
    \cos\varphi=\frac{{\mathbf v}\cdot {\mathbf m}}{|{\mathbf v}||{\mathbf m}|},
\end{equation}
where $\mathbf v$ is an instant vector of the fragment velocity and $\mathbf m$ is the vector towards the position of the final merging event, defined by
\begin{equation}
    {\mathbf m}={\mathbf r}_{\rm m}-{\mathbf r}.
\end{equation}
Here, $\mathbf r$ is an instantaneous position of the fragment and ${\mathbf r}_{\rm m}$ is the position of the final merging event. 

\begin{figure}
    \centering
    \includegraphics[width=0.46\textwidth]{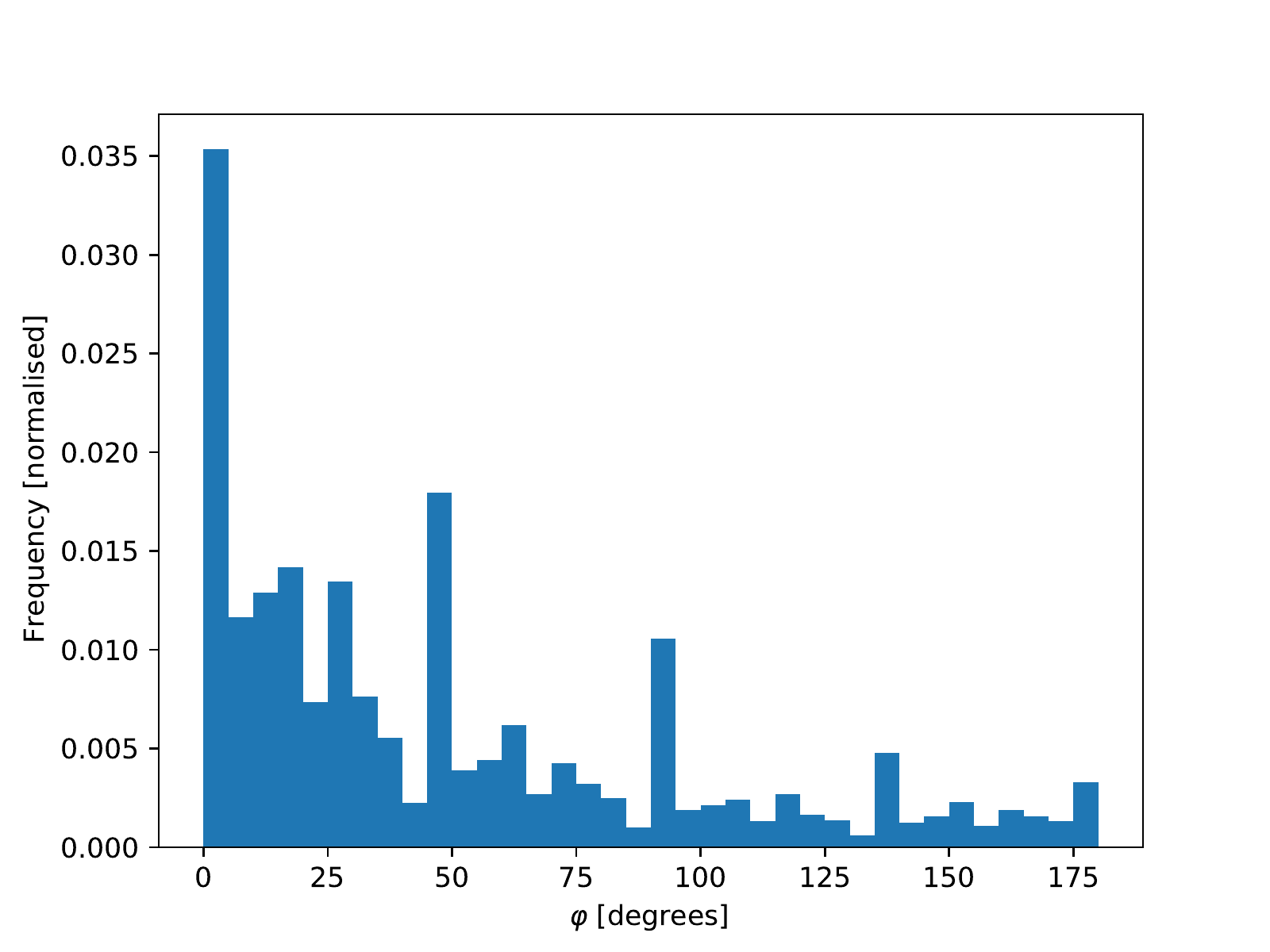}
    \caption{Histogram of the fragment's directions towards the merging centres.}
    \label{fig:hist_directions}
\end{figure}

The histogram of $\varphi$ displayed in Fig.~\ref{fig:hist_directions} indicates that, in general, fragments steer towards the merging position, but not quite straight. The variations are large, similar to a random-walk process. The peaks at multiples of 90 degrees are a consequence of the data sampling; a better spatial resolution would likely remove these peaks. 

\section{Conclusions}

We studied the critical evolution phases of more than 1000 active regions observed by the HMI/SDO instrument. The sample contained active regions both undergoing formation and decaying. Using a feature-recognition and tracking code, we automatically detected dark features representing magnetised features from small elements to proper sunspots. We tracked these objects through the spatio-temporal datacube and followed their evolution. We recorded events when these objects merged with other object and events when the object split in two. We studied the statistics of these events. 

The lifetime of the detected objects is rather short, typically a few tens of minutes. For large objects, the determination of the lifetime may be biased by our methodology. They merge with other objects or fade soon after their appearance. Their typical size seems to be around 2~Mm$^2$. Due to the short lifetime, the number of the appearing (by the emergence of splitting) objects and disappearing (by fading or merging) objects is in a surprising balance throughout the typical lifetime of the active region.  

We obtained our result by automatic processing of the large sample of active regions. Statistically, we confirm that sunspots form by merging events of smaller fragments. The coalescence process is driven by the turbulent diffusion in a process similar to random-walk, where supergranular flows seem to play an important role. The number of the appearing fragments does not seem to significantly correlate with the number of sunspots that formed. The formation seems to be consistent with the magnetic field accumulation. Statistically, the merging occurs between a large and a much smaller object. The decay of the active region seems to take place preferably by a process similar to the erosion. 

Both merging and splitting events preferably occur in the immediate vicinity of evolved sunspots. Merging or splitting further away is rare. The merging and splitting events do not occur at random places, the preferential locations cluster again in a close vicinity to the sunspots. 

Using HMI/SDO intensitygrams, we are unable to assess the scenario by \cite{1987SoPh..112...49G}. After a fragment merges into a sunspot, the track of its location and evolution is lost mainly due to the small spatial resolution of 1". With HMI/SDO intensitygrams, it is not possible to identify and track the dark nuclei within sunspot umbrae, which are seen in high-resolution observations \citep[e.g.][]{2007msfa.conf..205S}. These dark nuclei probably correspond to the surviving fragments indicated by the scenario by \citeauthor{1987SoPh..112...49G}.
Higher resolution observations perhaps with a better dynamical range are needed. This issue thus remains open.  

\begin{acknowledgements}
    The authors were supported by the Czech Science Foundation under the grant project 18-06319S. ASU CAS is funded by the institute research project ASU:67985815. We thank the referee for helpful suggestions for improvements of this paper. The SDO/HMI data are available by courtesy of NASA/SDO and the HMI science team.
\end{acknowledgements}

\end{document}